  \documentclass[journal,twocolumn]{IEEEtran}
\usepackage{adjustbox}
\usepackage{amsmath,amssymb,amsfonts}
\usepackage{colortbl}
\usepackage[table,xcdraw]{xcolor}
\usepackage{balance}
\usepackage{graphicx}
\usepackage{caption}
\usepackage [noadjust]{cite} 
\usepackage{bm}  
\usepackage{amsmath}
\usepackage{amssymb}
\usepackage{enumerate}
\usepackage{stfloats}
\usepackage{cases}
\usepackage{epstopdf}
\usepackage{soul,color} 
\usepackage{hyperref}
\usepackage[percent]{overpic}
\usepackage{tabularx}
\usepackage[table]{xcolor}
\usepackage{multirow}
\usepackage{booktabs}  

\usepackage{algorithm}

\begin{document}
	\raggedbottom
	\allowdisplaybreaks
        \title{
    Strategic Utilization of Cellular Operator Energy Storages for Smart Grid Frequency Regulation}
	\author{Narges Gholipoor, Farid Hamzeh Aghdam, Mehdi Rasti 
		\thanks{This paper is supported by Business Finland via REEVA project with
Grant Number: 10284/31/2022.  Also, it was supported from ”The University
of Oulu \& The Academy of Finland Profi6 336449”.}
\vspace{-0.3cm}
	}

	\maketitle
\begin{abstract}
The innovative use of cellular operator energy storage enhances smart grid resilience and efficiency. Traditionally used to ensure uninterrupted operation of cellular base stations (BSs) during grid outages, these storages can now dynamically participate in the energy flexibility market. This dual utilization enhances the economic viability of BS storage systems and supports sustainable energy management.
In this paper, we explore the potential of BS storages for supporting grid ancillary services by allocating a portion of their capacity while ensuring Ultra Reliable Low Latency (URLLC) requirements, such as meeting delay and reliability requirements. This includes feeding BS stored energy back into the grid during high-demand periods or powering BSs to regulate grid frequency. We investigate the impacts of URLLC requirements on grid frequency regulation, formulating a joint resource allocation problem. This problem maximizes total revenues of cellular networks, considering both the total sum rate in the communication network and BS storages’ participation in frequency regulation, while considering battery aging and cycling constraints.
Simulation results show that a network with 1500 BSs can increase power vacancy compensation from 31\% to 46\% by reducing reliability from $10^{-8}$ to $10^{-3}$. For a power vacancy of $-30$ MW, this varies from $9.3$ MW to $13.5$ MW, exceeding a wind turbine’s capacity.
\end{abstract}
	 \vspace{-0.3cm}

	
\thispagestyle{empty}

\vspace{-0.2cm}
\section{Introduction}
Achieving net-zero emissions by 2050 is one of the most significant United Nations (UN) Sustainable Development Goals (SDGs). Consequently, sustainability has become a crucial objective across various industries, including telecommunication and smart grid. To meet this goal, industries must manage their energy usage, enhance their Energy Efficiency (EE), and minimize environmental impacts \cite{ITU-TechnicalSmart,SWGIMT2030}. 
Maintaining a balance between demand and supply is crucial for the reliable and sustainable operation of  smart grid. An imbalance can result in large-scale blackouts \cite{6466419}. Variations in demand and supply lead to frequency fluctuations within the smart grid, making frequency a key indicator for evaluating both the balance between demand and supply and the overall reliability of the smart grid \cite{ding2023strategy}.
When there is extra energy in the smart grid, the frequency increases. Conversely, when the demand for energy exceeds the supply, the frequency decreases. To maintain the operation of a power system, the frequency must be kept within a specific range, which requires careful management of electricity supply/demand balance. Conventional power systems, with presence of rotating power plants, usually have high inertia which helps them to overcome the demand/supply mismatch easily. 

The electrical power system’s inertia has decreased due to the rise of Renewable Energy Sources (RESs) and their significant contribution to power generation. The reason is that RESs are highly intermittent, uncertain and sensitive to weather and climate conditions so they may be unavailable when needed. As a result, generation resources have to produce extra power which might force them into inefficient operation in order for the power system frequency to stay within an acceptable range. Besides, heavy reliance on conventional generating plants is costly and environmentally harmful. Thus, novel approaches are needed in order to improve the operational and dynamic stability of power systems with large shares of renewables. In low-inertia systems, different control methods like virtual inertia, DFIG-based wind turbines, Battery Energy Storage Systems (BESSs), and demand response can also help regulate frequency \cite{hosseini2022battery}. This recently witnessed accomplishment by a hundred megawatts BESSs installed at South Australia proves that BESSs are highly effective for Primary Frequency Control (PFC) with their high rate of response. In some European systems, BESSs have already become part of PFC services while UK National Grid now demands “enhanced frequency response,” which requires one second or less duration for the response time \cite{arrigo2020assessment}.

The ICT sector accounts for about 4.7\% of the world’s electrical energy and 1.7\% of CO$_2$ emissions \cite{suarez2012overview}. A major contributor to this consumption is the operation of Base Stations (BSs) in cellular networks. BSs consume more than 57\% of the total energy used by cellular networks \cite{9260159}. BS electricity consumption  includes dynamic and static components, with major energy demands arising from the dynamic part (communication units), which consists of  Power Amplifier (PA), Base Band Unit (BBU), and Transceiver (RF).  80\% of the total electricity consumption in BSs is due to communication units \cite{ding2023strategy}.  
On the other hand, with the anticipated surge in mobile data traffic and subscriber count by 2030, the increased deployment of BSs
will be necessary. This will result in an expected escalation in electricity consumption. 

Therefore, BSs in cellular networks represent a significant electricity demand. Equipped with battery storage, these BSs can supply energy back to the grid to aid in frequency regulating. When grid frequency dips, the stored energy in BSs can be discharged into the smart grid to elevate the frequency. Conversely, BSs draw electricity for their operation, and as traffic demand intensifies, so does their energy consumption. During peak hours, this heightened demand can precipitate outages. To circumvent frequency variations in the smart grid, BSs can utilize their battery storage to sustain their operations and also feed extra energy back to the smart grid to stabilize it, while maintaining communication requirements such as reliability and latency.
The unique position of cellular operators enables the leveraging of existing infrastructure to provide a dual service—maintaining network stability and contributing to smart grid flexibility. This dual utilization not only enhances the economic viability of the storage systems but also supports the broader goal of sustainable energy management.

Recently, \cite{9260159,6831472,6883952,6687957} have investigated BSs of cellular networks powered by smart grids and renewable energy, primarily focusing on minimizing the energy costs consumed by BS operation, without considering frequency regulation. In \cite{6831472,6883952,6687957}, the optimization problems are limited to energy management, neglecting the communication network and user requirements. In \cite{9260159}, an integrated approach is adopted to optimize both  the energy in smart grid  and the total bandwidth consumed in the cellular network, while ensuring that data rate requirements for cellular users are met. Moreover, in \cite{ding2023strategy}, the frequency regulation through BS battery storage is investigated by considering BSs as a black box and ignoring the communication requirements. 

In cellular networks, providing services for users and meeting their communication requirements necessitates the consumption of  transmit power which directly impacts electricity usage. 
As communication requirements become more stringent, such as when lower latency is required, more transmit power and consequently more electricity, is consumed. Therefore, when BS battery storage is utilized for smart grid ancillary services, it is crucial to keep a balance the revenues earned from participating in the smart grid ancillary services with the need to provide reliable communication services for users. 
This balance is particularly important under stringent conditions in the smart grid, such as frequency depletion, or in the communication network, such as low latency requirements.
However, considering frequency regulation and modeling the process of feeding energy back into the smart grid through BS battery storage for regulating frequency in the smart grid while guaranteeing all users' communication requirements, such as reliability and delay, remains an open problem. Hence, in this paper, we propose a resource allocation framework in which the smart grid and cellular communication network jointly regulate frequency in the smart grid while guaranteeing Ultra Reliable and Low Latency (URLLC) communication users' requirements for the first time. 
The contributions of this paper are as follows, many of which have been considered for the first time:
\begin{itemize}
\item We propose a resource allocation framework 
aims to regulate frequency in the smart grid through BS battery storages while ensuring URLLC users requirements in cellular network.
The framework outlines strategies for different load conditions in the smart grid, aiming to maintain grid stability while efficiently utilizing available electricity.
\item We analyze the impacts of communication Key Performance Indicators (KPIs) on the smart grid's electricity consumption and frequency regulation. The findings can guide cellular operators in managing their electricity consumption based on user requirements and the network specifications provided to users.
\item Our simulations demonstrate that by adjusting reliability and delay requirements, significant energy savings can be achieved. This saved energy can be utilized during peak load conditions, contributing to the frequency regulation of the smart grid. For instance, by reducing reliability from $10^{-8}$ to $10^{-3}$ or increasing delay from $0.5$ ms to $3$ ms, we can save around $13.5$ Mw in an area with $1500$ BSs.  This saved energy is equivalent to the capacity of a wind turbine. 
\end{itemize}

The rest of this paper is organized as follows. In Section II, the system model is described. In Section III, we formulate the optimization problem. Numerical results and simulations are presented in Section IV. Finally, Section V concludes the paper.
\vspace{-0.4cm}
\section{System Model} \label{Systemmodel}

\subsection{General Description}
In this system model, our aim is to analyze the impacts of communication KPIs on the  smart grid frequency regulation. 
Here, we consider a multi-cell network comprising several BSs, represented by \(\mathcal{J} = \{1, ..., J\}\), each equipped with a battery storage capacity of \(B_j\) kWh. All BSs are interconnected with the smart grid via fiber links. The battery storage of BSs serves dual purposes: 1) acting as a backup supply during outages for the communication network, and 2) functioning as a virtual power plant for smart grid frequency regulation.

In this section, we initially outline the time scale for the operation of our system model. Subsequently, we delve into the specifics of cellular communication and user requirements. We then model the electricity consumption and formulate the battery storage of BSs as a power supply. Lastly, we formulate the regulation of frequency in the smart grid through cellular network battery sto.

   \vspace{-0.4cm}
\subsection{Time Scale for Smart Grid and communication Network Operation}
\begin{figure}[t]
    \centering
    \includegraphics[width=0.95\linewidth]{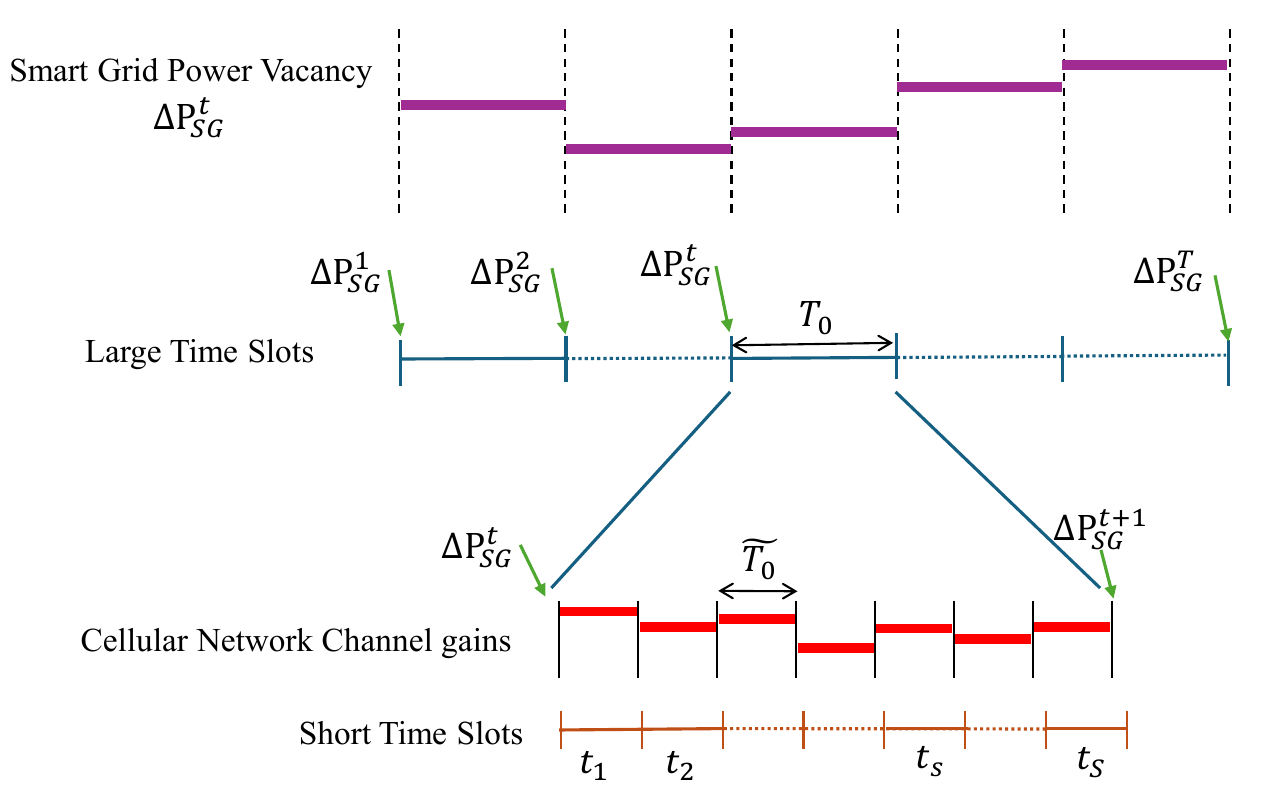}
    \caption{Network operation time scales}
    \vspace{-0.5cm}
    \label{fig4}
\end{figure}

The variation in wireless cellular channels and communication load occurs more rapidly than the frequency regulation decisions in the smart grid.  Therefore, similar to \cite{9260159}, we assume two time slots in our system model: a long time slot for frequency regulation decisions (on the order of 100 ms) \cite{heylen2021challenges}, and a short time slot (on the order of 10 ms) for the communication network. During the long time slot, the amount of power vacancy in the smart grid remains constant and  is transmitted to the communication network, which then communication network optimizes its power consumption accordingly. Moreover, we assume that during the short time slot, the channel gains are constant and vary from one short time slot to another. Therefore, each long time slot, denoted by $t$, with duration $T_0$ is divided into $S$ short time slots with duration $\tilde{T}_0 = \frac{T_0}{S}$, denoted by ${t_s}$, where $s$ represents the $s^{\text{th}}$ short time slot in long time slot $t$.
The set of short time slots in each long time slot denoted by \(\mathcal{S} = \{1, ..., S\}\). 
Hereinafter, $t$ signifies the long time slot, while $t_s$ represents the $s^{\text{th}}$ short time slot within long time slot $t$.
The time scale for operation of our proposed system are illustrated in Fig \ref{fig4}.
   \vspace{-0.3cm}
\subsection{Communication Network Model}
In the proposed system model, we consider a set of users, denoted by $\boldsymbol{\mathcal{{U}}_j}=\{1,\dots,U_j\}$, at BS $j$. The total number of users in our system model is represented by $\mathcal{U}_\text{Tot}=\bigcup_{j\in \mathcal{J}}\mathcal{U}_j$.  
Moreover, we consider $\boldsymbol{\mathcal{M}}=\{1,\dots,M\}$ as the set of all subcarriers for DownLink (DL) transmission between BSs and users.
Additionally, within our system model we consider specific user requirements denoted by $(R^{\text{TH}},  D^{\text{max}}, \epsilon^{{\rm{max}}})$,  where $R^{\text{TH}}$  represents the DL data rate requirement, $D^{\text{max}}$ signifies the maximum allowed End-to-End (E2E)  delay, and $\epsilon^{{\rm{max}}}$  represents the maximum tolerable E2E reliability, respectively.
It is worth noting that we adopt Orthogonal Frequency Division Multiple Access (OFDMA), allocating each subcarrier to at most one user per short time slot ${t_s}$.
	We define a binary variable $\rho_{u_{j}}^{m,t_s}$, which equals 1 if subcarrier $m$ is assigned to user $u_{j}$ at BS $j$ during short time slot ${t_s}$, and $0$ otherwise. 
Given that  we employ OFDMA in this setup, the following constraint ensures that each subcarrier is assigned to at most one user at each short time slot ${t_s}$:
	\begin{equation} \label{eqo2}
	\text{\text{C1: }}%
 \sum_{u_{j} \in \mathcal{U}_j} \rho_{u_{j}}^{m,t_s} \le 1, \forall j\in\mathcal{J}, m\in\mathcal{M},  s\in\mathcal{S}.\nonumber
	\end{equation}

\vspace{-0.15cm}
As mentioned earlier, in this work we address URLLC services. Thus, we employ the short block-length data rate formulation similar to  \cite{8399832, 8253477}. Consequently, the achievable rate for user $u_{j}$ over subcarrier $m$ at BS $j$ at short time slot ${t_s}$ is given by:
		\begin{align} \label{eqo31}
		&r_{u_{j}}^{m,t_s}= 
   \frac{{w^{{m,t_s}}{}}}{{\ln 2}} \bigg [ {\ln (1 + \gamma _{u_{j}}^{m,t_s}) - \sqrt {\frac{{\nu_{u_{j}}^{m,t_s}}}{\tilde{T}_0{w^{{m,t_s}}{}}}} Q^{ - 1}(\epsilon_{u_{j}}^{m,t_s,\text{DE}})} \bigg ],\nonumber\\ &\forall u_{j} \in \mathcal{U}_{j},  j \in \mathcal{J}, m \in \mathcal{M}, s \in \mathcal{S},
		\end{align}
where $w^{m,t_s}$ indicates the bandwidth of subcarrier $m$ at short time slot ${t_s}$ and $Q^{-1}(.)$ represents the inverse of the Q-function. Additionally, the decoding error probability for users $u_{j}$ over subcarrier $m$ at short time slot ${t_s}$ is represented by $\epsilon_{u_{j}}^{m,t_s,\text{DE}}$, and $\tilde{T}_0$ is the transmission duration, i.e., the time unit. Moreover, the channel dispersion, $\nu_{u_{j}}^{m,t_s}$, is defined as $\nu_{u_{j}}^{m,t_s}=1-\frac{1}{(1+\gamma^{m,t_s}_{u_{j}})^2}$ and can be approximated to 1 according to \cite{7529226,8541123}.
The Signal to Interference Plus Noise Ratio (SINR) for user $u_{j}$ on subcarrier $m$ at short time slot ${t_s}$, $\gamma^{m,t_s}_{u_{j}}$, is given by $\gamma^{m,t_s}_{u_{j}}=\frac{\rho_{u_{j}}^{m,t_s} p_{u_{j}}^{m,t_s}h_{u_{j}}^{m,t_s}}{\sigma_{u_{j}}^{m,t_s}+I_{u_{j}}^{m,t_s}}$, where $p_{u_{j}}^{m,t_s}$, $h_{u_{j}}^{m,t_s}$, and $\sigma_{u_{j}}^{m,t_s}$ denote the transmit power, channel power gain, and noise power from BS $j$ to user $u_{j}$ on subcarrier $m$ at short time slot ${t_s}$, respectively. The Inter-Cell Interference (ICI) for user $u_{j}$ on subcarrier $m$ at short time slot ${t_s}$, denoted as $I_{u_{j}}^{m,t_s}$, is calculated as  $I_{u_j}^{m,t_s} =\sum\limits_{\scriptstyle {j'} \in {\cal J}, j' \ne j 
}
{ {\sum\limits_{u_{j'} \in {\cal U}_{j'}} } } \rho_{u_{j'}}^{m,t_s}  p_{u_{j'}}^{m,t_s} h_{u_{j'},j}^{m,t_s},  \forall u_{j} \in \mathcal{U}_{j},    j \in \mathcal{J}, m \in \mathcal{M}, s \in \mathcal{S},$
where $h_{u_{j'},j}^{m,t_s}$ represents the channel power gain between BS $j$ and  user $u_{j'}$ over subcarrier $m$ at short time slot ${t_s}$. By considering  Eq. \eqref{eqo31}, the decoding error probability can be expressed as:
		\begin{align} \label{eqo311}
		&\epsilon_{u_{j}}^{m,t_s,\text{DE}}=  {Q} \bigg({\sqrt {\frac{{w^{{m,t_s}}{}}}{{\nu_{u_{j}}^{m,t_s}}}} \left[ {\ln (1 + \gamma _{u_{j}}^{m,t_s}) - \frac{{\Psi_{u_{j}}^{m,t_s}\ln 2 }}{{w^{m,t_s}\tilde{T}_0{}}}} \right]}\bigg), \nonumber\\ &\forall u_{j} \in \mathcal{U}_{j},  j \in \mathcal{J}, m \in \mathcal{M}, s \in \mathcal{S}, 
		\end{align}
where $\Psi_{u_{j}}^{m,t_s}=r_{u_{j}}^{m,t_s}.\tilde{T}_0$ represents the total number of transmitted bits in each short time unit, i.e., $\tilde{T}_0$, on subcarrier $m$ at short time slot ${t_s}$. For simplicity, we introduce     $\mu^{m,t_s}=w^{m,t_s}.\tilde{T}_0{}$, $A^{m}_{t_s}= \frac{1}{{2\pi \sqrt {{2^{2(\Psi_{u_{j}}^{m,t_s} /(\mu^{m,t_s} ))}} - 1} }}\sqrt{\mu^{m,t_s}}$, and $D^{m}_{t_s}= {2^{\Psi_{u_{j}}^{m,t_s} /(\mu^{m,t_s})}} - 1$. Given the lack of a closed-form expression for the Q-function, we approximate $\Xi(\gamma _{u_{j}}^{m,t_s})$  as $\Xi(\gamma _{u_{j}}^{m,t_s})\approx {Q}(\frac{{\ln (1 + \gamma_{u_{j}}^{m,t_s}) - {\rm{ \Psi_{u_{j}}^{m,t_s} /(}}\mu^{m,t_s} )}}{{\sqrt {\nu_{u_{j}}^{m,t_s}{{(\ln 2)}^2}/\mu^{m,t_s} } }})$ \cite{8399832}, which can be expressed as: 
		\begin{align} \label{eqQ-func}
		&	\Xi (\gamma _{u_{j}}^{m,t_s}) = \\& \left\{ {\begin{array}{*{20}{l}}
			{1,}&{\gamma _{u_{j}}^{m,t_s} \le D^{m}_{t_s} - \frac{1}{{2A^{m}_{t_s} }}},\\
			{1/2 - A^{m}_{t_s} (\gamma _{u_{j}}^{m,t_s} - {D^{m}_{t_s}}),}&{D^{m}_{t_s} - \frac{1}{{2 A^{m}_{t_s} }} \le \gamma_{u_{j}}^{m,t_s}   }\\
   &\le D^{m}_{t_s} + \frac{1}{{2A^{m}_{t_s} }},\\
			0,&{Z^{m,t_s}} + \frac{1}{{2A^{m}_{t_s} }} \le \gamma _{u_{j}}^{m,t_s}.
			\end{array}} \right.\nonumber
		\end{align}

 \vspace{-0.3cm} 
\subsection{Communication Network Constraints:}

Here, we articulate the communication requirements for users, including data rate, delay, and reliability.
  \subsubsection{Data Rate Requirement} The data rate requirement is encapsulated by the following constraint, which is derived from the service specifications:
		\begin{equation} \label{eqo3111}
		\begin{split}
		\text{C2: } r_{u_{j}}^{t_s}=\sum_{{m}\in \mathcal{M}} r_{u_{j}}^{m,t_s} \ge R^{\text{TH}}, \forall u_{j} \in \mathcal{U}_{j},  j \in \mathcal{J}, s \in \mathcal{S}, \nonumber
		\end{split}
		\end{equation}
where $R^{\text{TH}}$ denotes the minimum data rate requirement.
  
\subsubsection{Maximum Allowed E2E Delay}
The E2E delay in our system model consists of two components, 1) DL transmission delay  ($D_{u_{j}}^{{t_s},\text{T}}$), and 2) queuing delay at BS $j$ ($D_{u_{j}}^{{t_s},\text{Qu}}$). Due to the delay constraint for each user, we have
	\begin{align}
	&\text{C3: } D_{u_{j}}^{{t_s},\text{T}}+D_{u_{j}}^{{t_s},\text{Qu}}\le D^{{\rm{max}}}, \forall u_{j} \in \mathcal{U}_{j}, \nonumber  j \in \mathcal{J}, s \in \mathcal{S},  
	\end{align}
	where $D^{{\rm{max}}}$ is
	the maximum allowable E2E delay. 

  \textit{DL Transmission Delay}: To calculate the  DL transmission delays, we apply the following constraint:
	\begin{align} 
	&\text{C4: } D_{u_{j}}^{{t_s},\text{T}} \le \frac{C_{u_{j}}}{ r_{u_{j}}^{t_s}}, \forall u_{j} \in \mathcal{U}_{j},   j \in \mathcal{J}, s \in \mathcal{S},\nonumber
	\end{align}
	where $C_{u_{j}}$ represents the total transmitted bits of user $u_{j} $ at BS $j$.

 \textit{Queuing Delay:}
 We assume a First Come First Serve (FCFS) queue at each BS. The aggregation of arrival bits from all users at BS $j$ is modeled as a Poisson process \cite{9512400,9322582,1210731,8638940imp}. Consequently, the effective capacity for user $u_{j}$ at BS $j$ is determined as: 
	\begin{equation} \nonumber
	EC_{u_{j}}^{t_s}=-\Lambda_{u_{j}}^{t_s} \frac{(e^{-\theta_{u_{j}}^{t_s} }-1)}{\theta_{u_{j}}^{t_s}}, \forall u_{j} \in \mathcal{U}_{j},  j \in \mathcal{J}, s \in \mathcal{S}, \vspace{-0.2cm} 
	\end{equation}
	where $\theta_{u_{j}}^{t_s} \geq 0$ represents the statistical QoS exponent for user $u_{j}$ at BS $j$ at short time slot ${t_s}$. A larger  value of $\theta_{u_{j}}^{t_s}$ signifies a more stringent QoS requirement, while a smaller  value indicates a more relaxed QoS requirement. $\Lambda_{u_{j}}^{t_s}$ is defined the number of bits served per short time slot ${t_s}$ for user ${u_{j}}$  at BS $j$ queue, i.e., 
	${\Lambda_{u_{j}}^{t_s}} =  r_{u_{j}}^{t_s}$.
 Thus, 	the probability of queuing delay violation for user $u_{j}$ at BS $j$ queue at short time slot ${t_s}$ is determined as follows:
	\begin{align} \label{deqo17}
	&\Pr\{D_{u_{j}}^{{t_s},\text{Qu}}>D_{u_{j}}^{\text{Qu,Th}}\}=\vartheta  e^{(-\theta_{u_{j}}^{t_s} EC_{u_{j}}^{t_s} D_{u_{j}}^{\text{Qu,Th}})}\le {\epsilon_{u_{j}}^{{t_s},\text{Qu}}},\nonumber\\&  \forall u_{j} \in \mathcal{U}_{j},   j \in \mathcal{J}, s \in \mathcal{S}, 
	\end{align}
	where $D_{u_{j}}^{{t_s},\text{Qu}}$ represents the queuing delay for user $u_j$ at BS $j$ and $D_{u_{j}}^{\text{Qu,Th}}$ denotes the maximum allowed queuing delay. The parameter  $\vartheta  \le 1$ indicates the  probability of non-empty buffer, and $\epsilon_{u_{j}}^{{t_s},\text{Qu}}$ signifies the maximum allowable probability of queuing delay violation. Therefore, we have: 
	\begin{equation}\nonumber
	\begin{split}
	&\Pr\{D_{u_{j}}^{{t_s},\text{Qu}}>D_{u_{j}}^{\text{Qu,Th}}\} \leq e^ {( - {\theta_{u_{j}}^{t_s}}EC_{u_{j}}^{t_s} {D_{u_{j}}^{\text{Qu,Th}}})} =\nonumber\\& e^ {( {\theta_{u_{j}}^{t_s}}{\Lambda_{u_{j}}^{t_s}}\frac{{({e^{{-\theta_{u_{j}}^{t_s}}}} - 1)}}{{{\theta_{u_{j}}^{t_s}}}}{D_{u_{j}}^{\text{Qu,Th}}})}  =e^ {(- {\Lambda_{u_{j}}^{t_s}}(1-{e^{{-\theta_{u_{j}}^{t_s}}}}){D_{u_{j}}^{\text{Qu,Th}}})} \le \epsilon_{u_{j}}^{{t_s},\text{Qu}}.
	\end{split}
	\end{equation}
	Consequently, for E2E reliability we have the following constraint
	\begin{equation}
	\begin{split}
	{\text{C5: }}  r_{u_{j}}^{t_s}  \ge \frac{{\ln ({1/\epsilon_{u_{j}}^{{t_s},\text{Qu}}})}}{{(1- {e^{{-\theta_{u_{j}}^{t_s}}}}){D_{u_{j}}^{{t_s},\text{Qu}}}}}, \forall u_{j} \in \mathcal{U}_{j},  j \in \mathcal{J}, s \in \mathcal{S}.\nonumber
	\end{split}
	\end{equation}

  \subsubsection{Maximum Allowed E2E Reliability}
The E2E reliability  in our system model consists of two components, 1) DL decoding error probability  ($\epsilon_{u_{j}}^{{t_s},\text{DE}}$), and 2) queuing reliability at Bs $j$ ($\epsilon^{{t_s},\text{Qu}}$), i.e., delay violation probability. 
Due to the reliability constraint for each user, we have \cite{8541123}
	\begin{align}
	&\text{C6: } 1-(1-\epsilon_{u_{j}}^{{t_s},\text{DE}})(1-\epsilon_{u_{j}}^{{t_s},\text{Qu}})\approx \epsilon_{u_{j}}^{{t_s},\text{DE}}+\epsilon_{u_{j}}^{{t_s},\text{Qu}}\le \epsilon^{{\rm{max}}},\nonumber\\& \forall  u_{j} \in \mathcal{U}_{j},  j \in \mathcal{J},  s \in \mathcal{S},\nonumber
	\end{align}
	where $\epsilon^{{\rm{max}}}$ is
	the maximum allowable E2E reliability. 
\subsubsection{BS Transmit Power Constraint}
 According to \cite{lorincz2012measurements}, the transmit power of each BS is limited, leading to the following constraint:
 		\begin{equation} \label{equbsp}
 		\begin{split}
 	\text{C7: }  \sum_{u_{j} \in \mathcal{U}_{j}}\sum_{m \in \mathcal{M}} \rho_{u_{j}}^{m,t_s} p_{u_{j}}^{m,t_s} \le P^{\text{Max}}_j, \forall j \in \mathcal{J}, s \in \mathcal{S}, \nonumber 
 		\end{split}
 		\end{equation}
 where $P^{\text{Max}}_j$ is the maximum transmit power of BS $j$.
\vspace{-0.3cm}
\subsection{BS Electricity Consumption Model}
The total electricity consumption of BSs consists of two components: static and dynamic. The static component remains constant and does not depend on communication resource utilization, such as energy consumption for cooling, which depends on weather conditions. In contrast, the dynamic component varies with communication resource utilization. This relationship is modeled as a linear relation between communication resource utilization and electricity consumption by ETSI and ITU, the main standardization bodies \cite{ETSIBSPower,ITU-TPBS}. Therefore, the total electricity consumption of BS $j$ at short time slot ${t_s}$ is given by \cite{lorincz2012measurements,ETSIBSPower,ITU-TPBS}:
\begin{align}
  P^{{t_s}}_{j,\text{BS}}=U^{{t_s}}_{j,\text{BS}} P^{{t_s}}_{j,\text{Cof.}}+P^{{t_s}}_{j,\text{Static-BS}}, \forall j \in \mathcal{J},  
\end{align}
 where $P^{{t_s}}_{j,\text{Cof.}}$ is the electricity consumption  coefficient of resource utilization of BS $j$, i.e., $U^{{t_s}}_{j,\text{BS}}$. $U^{{t_s}}_{j,\text{BS}} P^{{t_s}}_{j,\text{Cof.}}$ and $P^{{t_s}}_{j,\text{Static-BS}}$ represent dynamic and static electricity consumption of BS $j$, respectively.
Therefore, the normalized of communication resource utilization is calculated as follows:
		\begin{equation} \label{equuuu}
		\begin{split}
	&
 U^{{t_s}}_{j,\text{BS}}=\frac{\sum_{m \in \mathcal{M}} \sum_{u_{j} \in \mathcal{U}_{j}}  
 \rho_{u_{j}}^{m,t_s} p_{u_{j}}^{m,t_s}}{P^{\text{Max}}_j},  \forall j \in \mathcal{J}. 
		\end{split}
		\end{equation}

\vspace{-0.4cm}
\subsection{BS's Battery Model}
As mentioned earlier, we assume that each BS is equipped with a battery with capacity $B_j$.
We define $y_{j}^{t} \in \{-1,0,+1\}$ to indicate the charging, discharging, and disconnected status of the battery storage of BS $j$ at long time slot $t$. If $y_{j}^{t}=1$, it signifies a charging status. If $y_{j}^{t}=-1$, it signifies a discharging status. If $y_{j}^{t}=0$, it means there is no need for charging or discharging. In other words, the BS battery is disconnected from the smart grid and is utilizing its own battery storage energy. In summery, we have:
	\begin{equation} \nonumber
	\begin{split}
	& y_{j}^{t}=\\&\begin{cases}
	-1, &\text{If the BS $j$ battery is discharging}\\ 
	0, & \text{If the BS $j$ battery is disconnected from the grid}\\
 +1&   \text{If the BS $j$ battery is charging}
	\end{cases}.
	\end{split}
	\end{equation}

Furthermore, we introduce $\kappa_{j}^{t}\in \{1,0\}$ as a binary variable, which denotes whether BS $j$ is directly utilizing the smart grid electricity or its battery storage for communication at long time slot $t$. If $\kappa_{j}^{t}=0$, it means BS $j$ is directly utilizing the smart grid electricity for communication. If $\kappa_{j}^{t}=1$, it means BS $j$ is directly utilizing energy from its battery storage for communication. In summery, we have:
	\begin{equation} \nonumber
	\begin{split}
	& \kappa_{j}^{t}=\begin{cases}
	+1, &\text{If the BS $j$ is utilized its battery as a supply}\\ 
	0, & \text{If the BS $j$ is utilized electricity as a supply}
	\end{cases}.
\end{split}
	\end{equation}
 
\begin{figure}
    \centering
    \includegraphics[width=.95\linewidth]{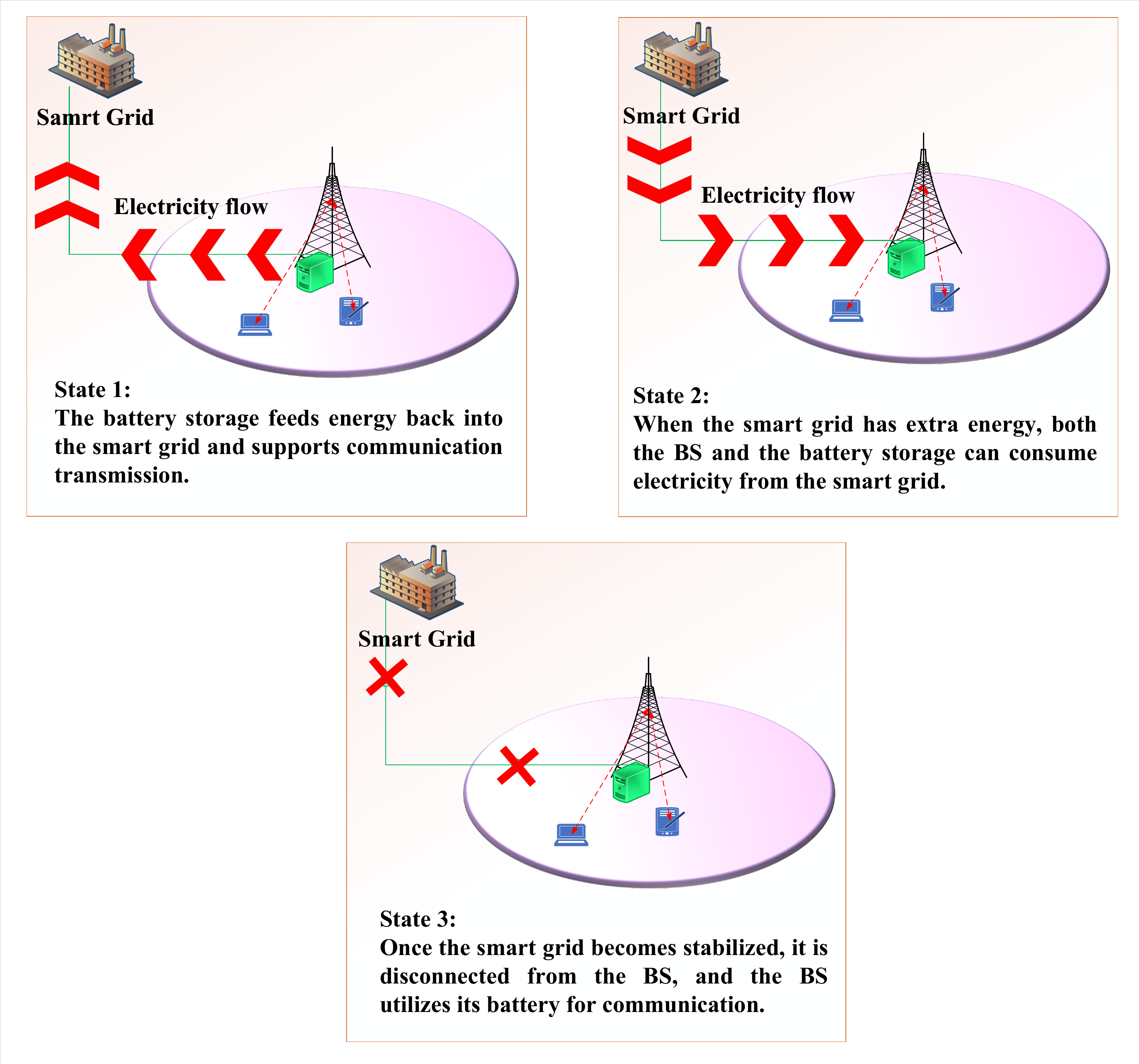}
    \caption{States of smart grid and BS battery storage}
    \vspace{-0.3cm}
    \label{fig2}
\end{figure}

As illustrated in Fig. \ref{fig2}, we define the three operational states of the BS battery storage with respect to energy management as follows:
\begin{enumerate}
    \item  When the smart grid requires additional energy to increase frequency, the BS battery storage should feed energy back into the smart grid. In this scenario, the BS consumes its own stored energy and also contributes energy from its battery storage back to the smart grid, i.e., $y_{j}^{t}=-1$,  $\kappa_{j}^{t}=1$.
    \item When the smart grid has excess energy, both the battery storage and the BS directly consume electricity from the smart grid to reduce frequency and achieve equilibrium. If the battery is fully charged, the BS alone consumes electricity for communication from the smart grid and the the battery storage is disconnected from the smart grid, i.e., $y_{j}^{t}=\{0,1\}$, $\kappa_{j}^{t}=0$.
    \item When the smart grid just becomes stabilized, the BS should be disconnected from the smart grid. Consequently, the BS relies on its battery storage to provide communication services, i.e., $y_{j}^{t}=0$, $\kappa_{j}^{t}=1$.
\end{enumerate}

The State of Charge (SoC) of BS $j$ battery storage  at long time slot $t$ is calculated as follows:
\begin{align}
    &{\text{SoC}}_{j}^{t}=(1-x_{j})\text{SoC}_{j}^{t-1}+ {T}_0 \eta_{j} y_{j}^{t} e_{j}^{t} -\sum_{s \in \mathcal{S}} \tilde{T}_0 \kappa_{j}^{t} F_{j}^{t_s} P^{{t_s}}_{j,\text{BS}},\nonumber
\end{align}
where  
$e_{j}^{t}=\frac{\tilde e_{j}^{t}}{B_j}$. Here, $\tilde e_{j}^{t}$  is the amount of electricity that BS $j$ receives/delivers from/to the smart grid at time slot $t$. 
$x_{j}$ is the discharging coefficient. 
$\eta_{j} \in [0,1]$ is the charging efficiency of BS $j$ battery storage. 
Furthermore, $F_{j}^{t_s}$ and $P^{{t_s}}_{j,\text{BS}}$ are consumption electricity coefficient and BS electricity consumption at short time slot ${t_s}$, respectively. 
It is worth noting that the SoC of BS $j$ battery storage and its variation should be maintained within a specific range to mitigate cycling aging \cite{9260159}.
Hence, we establish the following constraints \cite{9260159}:
\begin{align}
    \text{C8:~} \text{SoC}_{j}^{\text{Min}} \leq \text{SoC}_{j}^{t} \leq \text{SoC}_{j}^{\text{Max}}, \forall j \in \mathcal{J}, \nonumber 
\end{align}
\begin{align}
    \text{C9:~} \Delta\text{SoC}_{j}^{\text{Min}} \leq \text{SoC}_{j}^{t}-\text{SoC}_{j}^{t-1} \leq \Delta\text{SoC}_{j}^{\text{Max}}, \forall j \in \mathcal{J}, \nonumber
\end{align}
where $\text{SoC}_{j}^{\text{Min}}$ and $\text{SoC}_{j}^{\text{Max}}$ denote the minimum and maximum permissible SoC levels for BS $j$ battery storage. $\Delta\text{SoC}_{j}^{\text{Min}} $ and $\Delta\text{SoC}_{j}^{\text{Max}}$ are the minimum and maximum allowed  variation of SoC for BS $j$ battery storage.


\vspace{-0.3cm}
\subsection{Smart Grid Frequency Regulation Model}
In frequency regulation within the smart grid, it is crucial to balance the frequency and maintain it within a safe region by ensuring that demand and generated electricity are in equilibrium. If the frequency of the smart grid decreases, electricity must be injected to raise the frequency. Conversely, if the frequency increases, the excess electricity should either be consumed or electricity generation should be decreased to lower the frequency. The relationship between the power required to be injected into the smart grid or consumed, i.e., power vacancy, to balance the generated power and load demand is as follows:
		\begin{equation} \label{eqo21-fc}
		\begin{split}
		2 N \frac{\text{d}\Delta f}{\text{d}t}+ G. \Delta f= \frac{\Delta P_{\text{SG}}^{t}}{P_{\text{Base}}^{t}},
		\end{split}
		\end{equation}
where, 
$\Delta f= f^t-f^0$  denotes the frequency variation of smart grid, with $f^t$ and $f^0$  representing the smart grid frequency at long time slot $t$ and   nominal frequency, respectively.
Additionally, $N$, $G$, and $P_{\text{Base}}^{t}$ represent the inertia constant of the smart grid, the damping coefficient, and the base power, respectively.  $\Delta P_{\text{SG}}^{t}$  refers  to the power vacancy caused by the frequency deviation $\Delta f$,  which indicates the amount of power that needs to be injected into or consumed from the smart grid to regulate frequency.
Eq. \eqref{eqo21-fc} can be expressed as:
\begin{equation} \label{eqo21-1}
\begin{split}
2 N \frac{\text{d}{({f^t}-f^0)}}{\text{d}t}+ G. ({f^t-f^0})= \frac{\Delta P_{\text{SG}}^{t}}{P_{\text{Base}}^{t}}.
\end{split}
\end{equation}
By considering time slot duration $T_{0}$ and discretizing  Eq. \eqref{eqo21-fc},  we obtain:
\begin{equation} \label{eqo21-2}
\begin{split}
2 N \frac{{({f^t- f_{t-1}})}}{T_{0}}+ G. ({f^t-f^0})= \frac{\Delta P_{\text{SG}}^{t}}{P_{\text{Base}}^{t}}.
\end{split}
\end{equation}
  \vspace{-0.2cm}
Thus, $f^t$ is derived as follows:
\begin{equation} \label{eqo21-fccc}
\begin{split}
f^t=\frac{\frac{\Delta P_{\text{SG}}^{t}}{P_{\text{Base}}^{t}}+\frac{f^{t-1}}{T_0}2N+G f^0}{\frac{2 N}{T_0}+G}.
\end{split}
\end{equation}
As mentioned earlier, in frequency regulation, the frequency must be maintained within a safe operating range, $f^{\text{Min}}\leq f^t \leq f^{\text{Max}}$, where $f^{\text{Min}}$ and $f^{\text{Max}}$ represent the minimum and maximum allowable frequencies in the smart grid, respectively. 
It is important to note that the cellular network can not compensate all required power vacancy, i.e., $\Delta P_{\text{SG}}^{t}$, and just compensate a portion of that.
Therefore, the total consumed/injected power from/to  smart grid through cellular communication is obtained as follows:
\begin{align}
   &\sum_{s \in \mathcal{S}} \Delta P_{\text{Com}}^{{t_s}}=  \sum_{s \in \mathcal{S}}\sum_{j \in \mathcal{J}} y_{j}^{t}  e_{j}^{t} {T}_0 B_j+ (1-\kappa_{j}^{t})  P^{{t_s}}_{j,\text{BS}} \tilde{T}_0 B_j. \nonumber
\end{align}
The aim is to minimize the difference between $\Delta P_{\text{SG}}^{t}$ and $ \sum_{s \in \mathcal{S}} \Delta P_{\text{Com}}^{{t_s}}$, i.e., $|\Delta P_{\text{SG}}^{t} - \sum_{s \in \mathcal{S}} \Delta P_{\text{Com}}^{{t_s}}|$. 
Therefore, we rewrite this as follows:
\begin{align} \label{eqof}
   \text{C10:~}  z^{t} \Delta P_{\text{SG}}^{t}= \sum_{s \in \mathcal{S}} \Delta P_{\text{Com}}^{{t_s}},\nonumber 
\end{align}
where $0 \le z^{t} \le 1$ is the portion of the power vacancy, i.e., $\Delta P_{\text{SG}}^{t}$, of smart grid can be compensated through cellular BSs at long time slot $t$. 
Therefore, the aim is to maximize the participation of cellular network, i.e., $z^{t}$,  in frequency regulation in the smart grid. 
\vspace{-0.35cm}
\section{Optimization Problem Formulation}
In this paper, we define a multi-objective optimization problem aimed at maximizing the total revenue of a cellular network operator. This total revenue comprises the revenue earned from selling communication services to customers and the revenue obtained from the selling stored energy to the smart grid for frequency regulation. 
Therefore, we have
\begin{align}
 & \text{Rev}^{t} =\beta_\text{Cell} \tilde {\text{Rev}}_\text{Cell}^{t} + \beta_\text{Freq.} \text{Rev}_\text{Freq.}^{t}.
\end{align}
where, $\beta_\text{Cell}$ and $\beta_\text{Freq.}$ are coefficient of communication service revenue and smart grid frequency regulation revenue, respectively.
Moreover, smart grid frequency regulation revenue is $\text{Rev}_\text{Freq.}^{{t}} = z^{t}$. Moreover, the  communication service revenue at long time slot $t$ is obtained as follows:
  \begin{align}
 & \tilde {\text{Rev}}_\text{Cell}^{t}=\sum_{s \in \mathcal{S}}\sum_{j \in \mathcal{J}}  \sum_{u_{j} \in \mathcal{U}_{j}} r_{u_{j}}^{t_s}.
 \end{align}
The nature of these revenues is inherently inconsistent and distinct, making direct addition impractical. A straightforward summation of the two revenues without appropriate coefficients would disproportionately emphasize the revenue from communication services due to its dominant nature, resulting in a total rate exceeds 1, i.e., surpassing the maximum value of $z^t$.
Building upon the methodology outlined in \cite{9204689}, we introduce the normalization factor $a_0$ to balance the revenues, thereby ensuring an accurate representation of total revenue. Consequently, we have:
\begin{align}
 & \text{Rev}^{t} =a_0  \text{Rev}_\text{Cell}^{t} + (1-a_0) \text{Rev}_\text{Freq.}^{t}.
\end{align}
where $\text{Rev}_\text{Cell}^{t}=\tilde{a} \tilde {\text{Rev}}_\text{Cell}^{t}$ and $\tilde{a} = \frac{1}{R^{\text{Max}}}$, with $R^{\text{Max}}$ being the maximum total achievable rate in the cellular network.
We define $\beta_\text{Cell}=a_0 \tilde{a}$ and $\beta_\text{Freq.}=1-a_0$.  Moreover, $a_0$ is the coefficient to balance $\text{Rev}_\text{Cell}^{{t}}$ and $\text{Rev}_\text{Freq.}^{{t}}$, which means that if $a_0 = 0$, $\text{Rev}_\text{Freq.}^{{t}}$ will be optimized, and in the cellular network, only the constraints are satisfied. If $a_0 = 1$, $\text{Rev}_\text{Cell}^{{t}}$ will be optimized, and only a feasible point for frequency regulation is found. The optimal $a_0$ is determined in the simulation section.
Therefore, based on the mentioned constraints C1-C10, the optimization problem can be written as
	\begin{align}\label{optproblem}
	&\mathop {\max }\limits_{\scriptstyle{\bf{ P}},{\boldsymbol{\rho}}, {\bf{Y}}\hfill\atop
		\scriptstyle{\bf{Z}},{ \bf{E}},{\bf{K}}\hfill} \text{Rev}^{t},
	\\\text{s.t.:}
 &\text{\text{C1: }}\sum_{u_{j} \in \mathcal{U}_{j}} \rho_{u_{j}}^{m,t_s} \le 1, \forall j\in\mathcal{J}, m\in\mathcal{M}, \nonumber\\
 &\text{C2: } r_{u_{j}}^{t_s} \ge R^{\text{TH}}, \forall u_{j} \in \mathcal{U}_{j},  j \in \mathcal{J}, \nonumber\\
 &\text{C3: } 
 D_{u_{j}}^{{t_s},\text{T}}+D_{u_{j}}^{{t_s},\text{Qu}}\le D^{{\rm{max}}}, \forall u_{j} \in \mathcal{U}_{j},   j \in \mathcal{J},  \nonumber\\
 &\text{C4: } D_{u_{j}}^{{t_s},\text{T}} \le \frac{C_{u_{j}}}{ r_{u_{j}}^{t_s}}, \forall u_{j} \in \mathcal{U}_{j},   j \in \mathcal{J},
 \nonumber\\
&{\text{C5: }} r_{u_{j}}^{t_s}  \ge \frac{{\ln ({1/\epsilon_{u_{j}}^{{t_s},\text{Qu}}})}}{{(1- {e^{{-\theta_{u_{j}}^{t_s}}}}){D_{u_{j}}^{{t_s},\text{Qu}}}}}, \forall  u_{j} \in \mathcal{U}_{j},  j \in \mathcal{J}, \nonumber\\
 &\text{C6: }  \epsilon_{u_{j}}^{{t_s},\text{DE}}+\epsilon_{u_{j}}^{{t_s},\text{Qu}}\le \epsilon^{{\rm{max}}}, \forall  u_{j} \in \mathcal{U}_{j},  j \in \mathcal{J},  \nonumber\\
 & \text{C7: }  \sum_{u_{j} \in \mathcal{U}_{j}}\sum_{m\in \mathcal{M}}\rho_{u_{j}}^{m,t_s} p_{u_{j}}^{m,t_s} \le P^{\text{Max}}_j, \forall j \in \mathcal{J}, \nonumber\\
 & \text{C8:~} \text{SoC}_{j}^{\text{Min}} \leq \text{SoC}_{j}^{t} \leq \text{SoC}_{j}^{\text{Max}}, \forall j \in \mathcal{J}, \nonumber\\
  &  \text{C9:~} \Delta\text{SoC}_{j}^{\text{Min}} \leq \text{SoC}_{j}^{t}-\text{SoC}_{j}^{t-1} \leq \Delta\text{SoC}_{j}^{\text{Max}}, \forall j \in \mathcal{J}, \nonumber\\
  &\text{C10:~}  z^{t} \Delta P_{\text{SG}}^{t}=  \sum_{s \in \mathcal{S}} \Delta P_{\text{Com}}^{{t_s}}, \nonumber\\
   &\text{C11:~}  0 \leq z^{t} \leq 1. \nonumber
	\end{align}
The optimization variables in \eqref{optproblem} are  transmit power allocation, subcarrier allocation, smart grid electricity allocation, BSs battery status, type of power supply selection for BS, and power vacancy compensation, denoted by $\boldsymbol{P}$,   $\boldsymbol{\rho}$, $\boldsymbol{E}$,  $\boldsymbol{Y}$, $\boldsymbol{K}$, and  $\boldsymbol{Z}$, respectively. 
In problem \eqref{optproblem}, the data rate functions are non-convex, leading to the overall non-convexity of the problem. Additionally, this problem involves both discrete and continuous variables, further complicating its solution.  
Moreover, in problem \eqref{optproblem}, if the reliability values, i.e, $\epsilon_{u_{j}}^{{t_s},\text{DE}}$, $\epsilon_{u_{j}}^{{t_s},\text{Qu}}$,  are treated as optimization variable, this problem becomes intractable as stated in \cite{8253477}. Therefore,following the approach in  \cite{8253477}, we assume $ \epsilon_{u_{j}}^{{t_s},\text{DE}}=\epsilon_{u_{j}}^{{t_s},\text{Qu}}= \frac{\epsilon^{{\rm{max}}}}{2}$.
\section{Optimization Problem Solution}
The optimization problem, as defined in \eqref{optproblem}, is a mixed-integer non-convex program that presents significant challenges to solve. To facilitate the resolution of this problem, we introduce a new variable $\dot{e}_{j}^{t} = y_{j}^{t} e_{j}^{t}$. Furthermore, we define an auxiliary decision variable $\boldsymbol{D}$  for delay adjustment.
To address the optimization problem in \eqref{optproblem}, we adopt the Alternative Search Method (ASM), known to converge towards a suboptimal solution as referenced in \cite{6678362}. Based on the principles of the ASM approach, we decompose the problem into four subproblems: 1) Frequency regulation subproblem, 2) Subcarrier allocation subproblem, 3) Transmit power allocation subproblem, and 4) Delay adjustment subproblem. 
Initially, we identify a feasible point for the proposed problem, denoted as $\boldsymbol{\rho}[0]$, $\boldsymbol{P}[0]$, $\boldsymbol{E}[0]$, $\boldsymbol{Y}[0]$, $\boldsymbol{K}[0]$, and $\boldsymbol{Z}[0]$. At the beginning of each iteration and each long time slot, the frequency regulation subproblem is solved, and its output is considered fixed for the other subproblems. Then, during each short time slot, Phase 2 subproblems, i.e., subcarrier allocation subproblem, transmit power allocation subproblem, and delay adjustment subproblem, are solved iteratively.
It is worth noting that we optimize the target variable by considering the others as fixed values based on the previous iteration. 
It is worth noting that the iterations conclude when the objective functions remain unchanged or the number of iterations exceeds a predefined threshold.
In this section, we explore each subproblem and propose their respective solutions.

\vspace{-0.3cm}
 \subsection{Frequency Regulation Subproblem}
The frequency regulation subproblem with fixed assigned subcarrier and allocated transmit power  is written as follows:
	\begin{align}\label{electricity subproblem}
	&\mathop {\max }\limits_{\scriptstyle{\bf{}}{\bf{Z}},{\bf{}}\hfill\atop
		\scriptstyle{\bf{}}{ \bf{\hat E}},{\bf{K}}\hfill} \text{Rev}^{t},
	\\\text{s.t.:}&\text{(C8)-(C11)}. \nonumber
	\end{align}
This subproblem is formulated as a Mixed Integer Linear Programming (MILP) problem and can be solved using any existing optimization toolbox, such as Gurobi or MOSEK, in Matlab.
\vspace{-0.2cm}
\subsection{Subcarrier allocation subproblem}
Given allocated transmit power, selected communication power supply, allocated electricity, the subcarrier allocation subproblem,  is as follows:
	\begin{align}\label{Subcarrier Allocation}
	&\mathop {\max }\limits_{\scriptstyle{ \boldsymbol{\rho}}} \text{Rev}^{t},
	\\\text{s.t.:}&\text{~(C1)-(C2), (C4)-(C5), (C7)-(C10)}. \nonumber
	\end{align}
This problem is non-convex due to the data rate function. By relaxing $\rho_{u_{j}}^{m,t_s} \in [0,1]$ and employing the Successive Convex Approximation (SCA)-based Difference of Convex (DC) functions approach, similar to \cite{9146520,6678362}, it can be transformed into a convex problem. DC transformation for short block-length data rate 
is explained in detail in \cite{9146520}. Hence, we write the rate function as the difference of two concave functions as follows:
	\begin{align}\label{DCRate}
	& r_{u_{j}}^{t_s}= \sum_{m \in \mathcal{M}}  \log_2(1+\frac {\rho_{u_{j}}^{m,t_s} p_{u_{j}}^{m,t_s}h_{u_{j}}^{m,t_s}}{\sigma_{u_{j}}^{m,t_s}+ \sum\limits_{\scriptstyle {j'} \in {\cal J},\hfill\atop		\scriptstyle j' \ne j\hfill
}
{ {\sum\limits_{u_{j'} \in {\cal U}_{j'}} } } \rho_{u_{j'}}^{m,t_s}  p_{u_{j'}}^{m,t_s} h_{u_{j'},j}^{m,t_s}})\nonumber\\&=\sum_{m \in \mathcal{M}} \log_2(\sigma_{u_{j}}^{m,t_s}+\nonumber\\& \sum\limits_{\scriptstyle {j'} \in {\cal J},\hfill\atop		\scriptstyle j' \ne j\hfill
}
{ {\sum\limits_{u_{j'} \in {\cal U}_{j'}} } } \rho_{u_{j'}}^{m,t_s}  p_{u_{j'}}^{m,t_s} h_{u_{j'},j}^{m,t_s}+ {\rho_{u_{j}}^{m,t_s} p_{u_{j}}^{m,t_s}h_{u_{j}}^{m,t_s}})\nonumber\\&-  \log_2(\sigma_{u_{j}}^{m,t_s}+ \sum\limits_{\scriptstyle {j'} \in {\cal J},\hfill\atop		\scriptstyle j' \ne j\hfill
}
{ {\sum\limits_{u_{j'} \in {\cal U}_{j'}} } } \rho_{u_{j'}}^{m,t_s}  p_{u_{j'}}^{m,t_s} h_{u_{j'},j}^{m,t_s})\nonumber\\&=f(\boldsymbol{\rho})-g(\boldsymbol{\rho}),
	\end{align}
	where $f(\boldsymbol{\rho})$ and $g(\boldsymbol{\rho})$ are concave functions as follows:
	\begin{align}
	& f(\boldsymbol{\rho})=\sum_{m \in \mathcal{M}} \log_2(\sigma_{u_{j}}^{m,t_s}+\nonumber\\& \sum\limits_{\scriptstyle {j'} \in {\cal J},\hfill\atop		\scriptstyle j' \ne j\hfill
}
{ {\sum\limits_{u_{j'} \in {\cal U}_{j'}} } } \rho_{u_{j'}}^{m,t_s}  p_{u_{j'}}^{m,t_s} h_{u_{j'},j}^{m,t_s}+ {\rho_{u_{j}}^{m,t_s} p_{u_{j}}^{m,t_s}h_{u_{j}}^{m,t_s}}),\nonumber
	\end{align}
	\begin{equation}
	\begin{split}
	g(\boldsymbol{\rho})= \sum_{m \in \mathcal{M}} \log_2(\sigma_{u_{j}}^{m,t_s}+ \sum\limits_{\scriptstyle {j'} \in {\cal J},\hfill\atop		\scriptstyle j' \ne j\hfill
}
{ {\sum\limits_{u_{j'} \in {\cal U}_{j'}} } } \rho_{u_{j'}}^{m,t_s}  p_{u_{j'}}^{m,t_s} h_{u_{j'},j}^{m,t_s}).\nonumber
	\end{split}
	\end{equation}
By adopting DC approach to transform $f(\boldsymbol{\rho})-g(\boldsymbol{\rho})$ into a convex form,  we employ the following linear approximation for each  user $u_{j}$ based on first order Taylor series in the point ${\boldsymbol{\rho}}^l$ as follows:
	\begin{equation}
	\begin{split}
	g(\boldsymbol{\rho}^l)=g(\boldsymbol{\rho}^{l-1})+\nabla g(\boldsymbol{\rho}^{l-1})({{\boldsymbol{\rho}}^l}-{\boldsymbol{\rho}}^{l-1}),\nonumber
	\end{split}
	\end{equation}
	where $l$ indicates the iteration numbers and $\nabla g(\boldsymbol{\rho}^{l-1})$ is obtained as follows:
	\begin{equation}
	\begin{split}
	\nabla g(\boldsymbol{\rho}^{l-1})= \left\{ {\begin{array}{*{20}{c}}
		0,&{{\rm{if~~~}}j' = j},\\
		\frac{p_{u_j^s}^{m,t_s}h_{u_j^s}^{m,t_s}}{{\sigma _{u_j^s}^{m,t_s} + I_{u_j^s}^{m,t_s} }},&{{\rm{if~}}j' \ne j}.
		\end{array}} \right.\nonumber
	\end{split}
	\end{equation}
 Therefore the data rate function in Eq. \eqref{DCRate} is transformed into the convex function.
 \vspace{-0.2cm}
\subsection{Transmit Power Allocation Subproblem}
By assuming subcarrier assignment, communication power supply selection, electricity allocation, and delay as fixed variables, the transmit power allocation subproblem is formulated as follows:
	\begin{align}\label{Power allocation subproblem}
	&\mathop {\max }\limits_{\scriptstyle{ \bf{P}}} \text{Rev}^{t},
	\\\text{s.t.:}&\text{~(C2),(C4)-(C5), (C7)-(C10)}. \nonumber
	\end{align}
This problem is non-convex due to the data rate function; however,  it can be transformed into a convex problem using the DC approach, similar to the subcarrier allocation subproblem.

 \vspace{-0.3cm}
 \subsection{Delay Adjustment Subproblem}
The delay adjustment subproblem  is written as follows:
	\begin{align}\label{delay subproblem}
	&
 \text{find} (\bf{D}),
	\\\text{s.t.:}&\text{~(C3)-(C5)}. \nonumber
	\end{align}
This subproblem is formulated as a  Linear Programming (LP) problem and can be solved using any existing optimization toolbox, such as Gurobi or MOSEK, in Matlab.

\section{Computational Complexity}
In the ASM approach, the total computational complexity of the problem is calculated by multiplying the total number of iterations by the summation of the complexities of all subproblems, as stated in \cite{8686217}.
It is worth noting that here we calculate the maximum computational complexity, which occurs at the beginning of the large time slot.
Therefore, the total computational complexity is $O^{\text{Tot}} = L(O^1 + O^2 + O^3+O^4)$, 
where $L$ is the number of iterations, and $O^1$, $O^2$,  $O^3$, and $O^4$ are the computational complexities of the frequency regulation subproblem, subcarrier allocation subproblem, transmit power subproblem, and  delay adjustment subproblem, respectively. 
All subproblems are solved via the CVX toolbox in MATLAB, which exploits the Interior Point Method (IPM) to solve the problem \cite{grant2014cvx}. Thus, the computational complexity of each subproblem is calculated as follows:
\begin{align}
    O^i = \frac{\log \text{CO.}^i}{q^0 \varrho \log \tilde{\zeta}},
\end{align}
where $\text{CO}^i$ is the total number of constraints of subproblem $i$. Moreover, $\tilde{\zeta}$ is used for updating the accuracy, and $q^0$ and $\varrho$ represent the initial point for approximation and the stopping criteria of IPM, respectively. The values of $\text{CO}^i$ are calculated in Table \ref{table-C}.

	\begin{table}[]
		\centering
		\caption{Computational Complexity}
		\label{table-C}
  \begin{adjustbox}{width=0.49\textwidth}
		\begin{tabular}{|c|c|}
			\hline
			\textbf{Subproblems}             & \textbf{Computational Complexity}                                                                                                                                            \\ \hline
   Frequency Regulation subproblem & $\mathcal{O}\left(\dfrac{\log(2J+2)/q^0 \varrho}{\log \tilde\zeta}\right)$ \\ \hline
			Subcarrier Allocation subproblem & $\mathcal{O}\left(\dfrac{\log(MJ+3U_ \text{Tot}+3J+1)/q^0 \varrho}{\log \tilde\zeta}\right)$ \\ \hline
			Power Allocation subproblem      & $\mathcal{O}\left(\dfrac{\log(5U_ \text{Tot}+3J+1)/q^0 \varrho}{\log\tilde \zeta}\right)$                                                                                                \\ \hline
		 \begin{tabular}[c]{@{}c@{}}Delay Adjustment Subproblem \end{tabular}	     & $\mathcal{O}\left(\dfrac{\log(3U_ \text{Tot})/ q^0  \varrho}{\log\tilde  \zeta}\right)$                                                                             \\ \hline
		\end{tabular}
  		\end{adjustbox}
    \vspace{-0.3cm}
	\end{table}
\vspace{-0.4cm}
\section{Simulation Results}
	
	\begin{table}[t]
		\centering
		\caption{Considered parameters in numerical results}
		\label{table55}
			\begin{adjustbox}{width=0.49\textwidth}
		\begin{tabular}{|c|c|}
			\hline
			\textbf{Parameter}                          & \textbf{Value of each parameter}      \\ \hline
Number of BSs                               & 1500              \\ \hline
Number of Users per Cell                   \cite{9260159}            & 20              \\ \hline
Cell Radius           \cite{8399832}                   & $500$m              \\ \hline
			Path loss exponent   \cite{ngo2014joint}                       & $3$                                   \\ \hline
			QoS exponent              \cite{8070468}                  & $0.05$                                  \\ \hline
			PSD  of the received AWGN noise  \cite{ngo2014joint}           & $-174$ dBm/Hz                         \\ \hline
			Bandwidth of DL         \cite{8399832}                    & $20$~MHz                          \\ \hline
			Number of DL subcarriers                    & $32$                                  \\ \hline
			Maximum transmit power of the BS \cite{8253477,8399832}           & $46$~dBm                         \\ \hline
			Required Reliability \cite{8070468,7841709,8253477}  & $10^{-7}$                         \\ \hline
   E2E Delay \cite{8399832} & $1$ ms    \\ \hline
   			Packet Size \cite{8253477,8399832} & $20$ bytes                        \\ \hline
      Noise Spectral Density \cite{8253477} & $-174$ dBm/Hz                        \\ \hline
 			             BSs battery Capacity \cite{ding2023strategy}                  & $ 20$KWh \\ \hline
$\text{SoC}_{j}^{\text{Min}}$ \cite{ding2023strategy}                   & $0.3 $ \\ \hline
$\text{SoC}_{j}^{\text{Max}}$  \cite{ding2023strategy}                  & $0.9 $ \\ \hline
$ \Delta\text{SoC}_{j}^{\text{Min}}$    \cite{6883952}               & - 0.5 \\ \hline
$ \Delta\text{SoC}_{j}^{\text{Max}}$        \cite{6883952}            & + 0.5\\ \hline
$\Delta P_{\text{SG}}^{t}$      \cite{ding2023strategy}              & $\{-30,0,+30\}$ MW \\ \hline
		\end{tabular}
		\end{adjustbox}
  \vspace{-0.3cm}
	\end{table}

	\subsection{Simulation Setup}
	This section presents simulation results to assess the performance of the proposed system model. In our simulations, the BSs are uniformly located from each other. Furthermore, we adopt a Rayleigh fading wireless channel model, where the channel power gains of subcarriers are independent. The channel power gains for the radio access links are defined as $h_{u_{j}}^{m,t_s}=\Omega_{u_{j}}^{m,t_s}{d_{u_{j}}}^{\upsilon}$, with $d_{u_{j}}$ representing the distance between user $u_{j}$ and BS $j$, $\Omega_{u_{j}}^{m,t_s}$ following a Rayleigh distribution as a random variable, and $\upsilon=3$ denoting the path-loss exponent. The power spectral density (PSD) of the additive white Gaussian noise (AWGN) is set to $-174$ dBm/Hz. The system model parameters are summarized in Table \ref{table55}. From the smart grid perspective, it is assumed that the BSs are distributed in IEEE 14-bus test network. Further it is assumed that a sudden change of demand has been occurred in seconds 8 and 16 to observe the frequency response of the system	\cite{ngo2014joint,9851476,8399832,ding2023strategy,8253477,8070468,7841709}. 

 \vspace{-0.3cm}
\subsection{Optimum $a_0$}
\begin{figure}[t]
    \centering
    \includegraphics[width=0.93\linewidth]{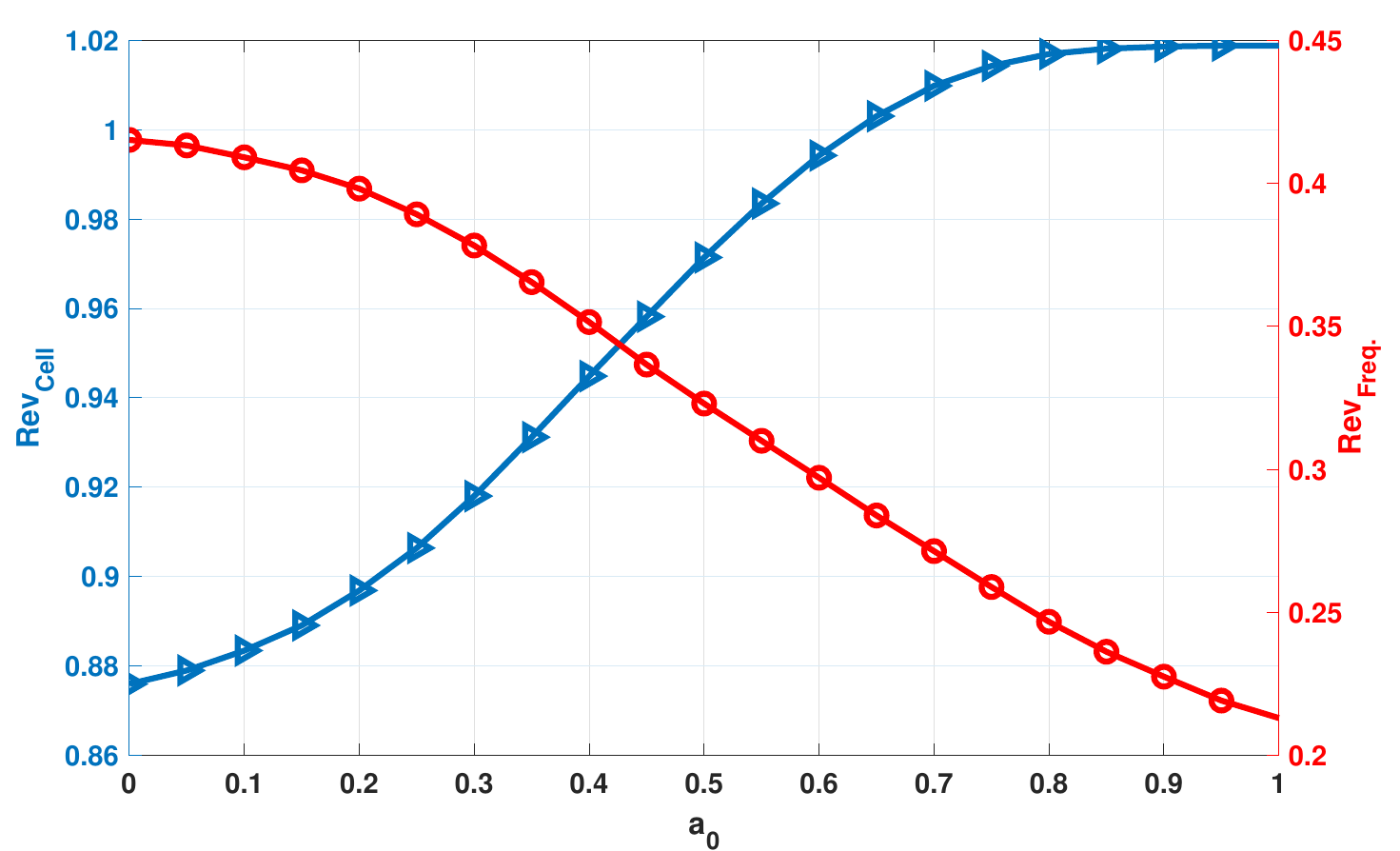}
    \caption{Impacts of weighted factor on the objectives}
    \vspace{-0.5cm}
    \label{SimFig-1}
\end{figure}

\begin{figure}[t]
    \centering
    \includegraphics[width=0.93\linewidth]{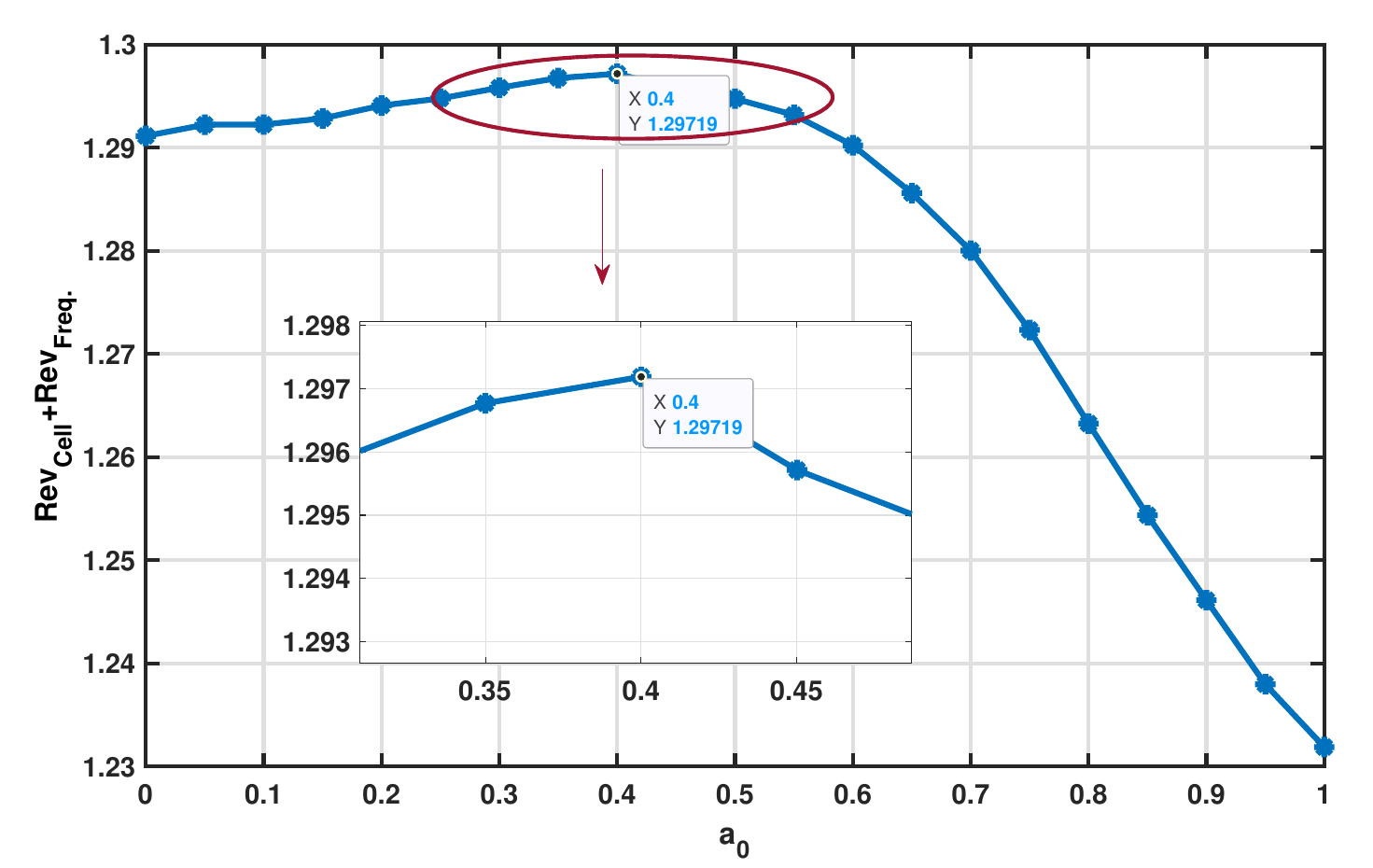}
    \caption{Impacts of weighted factor on the total revenues}
    \vspace{-0.6cm}
    \label{SimFig-1-1}
\end{figure}

The impact of $a_0$ on the revenues of the communication service ($\text{Rev}_\text{Cell}$) and participation in frequency regulation in the smart grid  ($\text{Rev}_\text{Freq.}$) is illustrated in Fig. \ref{SimFig-1}. As shown, by increasing $a_0$, $\text{Rev}_\text{Cell}$ increases while $\text{Rev}_\text{Freq.}$ decreases. For instance, when $a_0=0$, the revenue from smart grid frequency regulation is maximized, while the communication requirements for the cellular network are just satisfied. Conversely, when $a_0=1$, the communication service revenue is maximized, and the remaining power in battery capacity can participate in smart grid frequency regulation. Particularly, as illustrated in Fig. \ref{SimFig-1-1} when $a_0=0.40$, the total revenue reaches its maximum value. Hence, for the rest of the simulation section, we set $a_0$ to $0.40$.


  \vspace{-0.3cm}
\subsection{Communication KPIs Impacts on frequency Regulation}
\begin{figure}[t]
    \centering
    \includegraphics[width=0.93\linewidth]{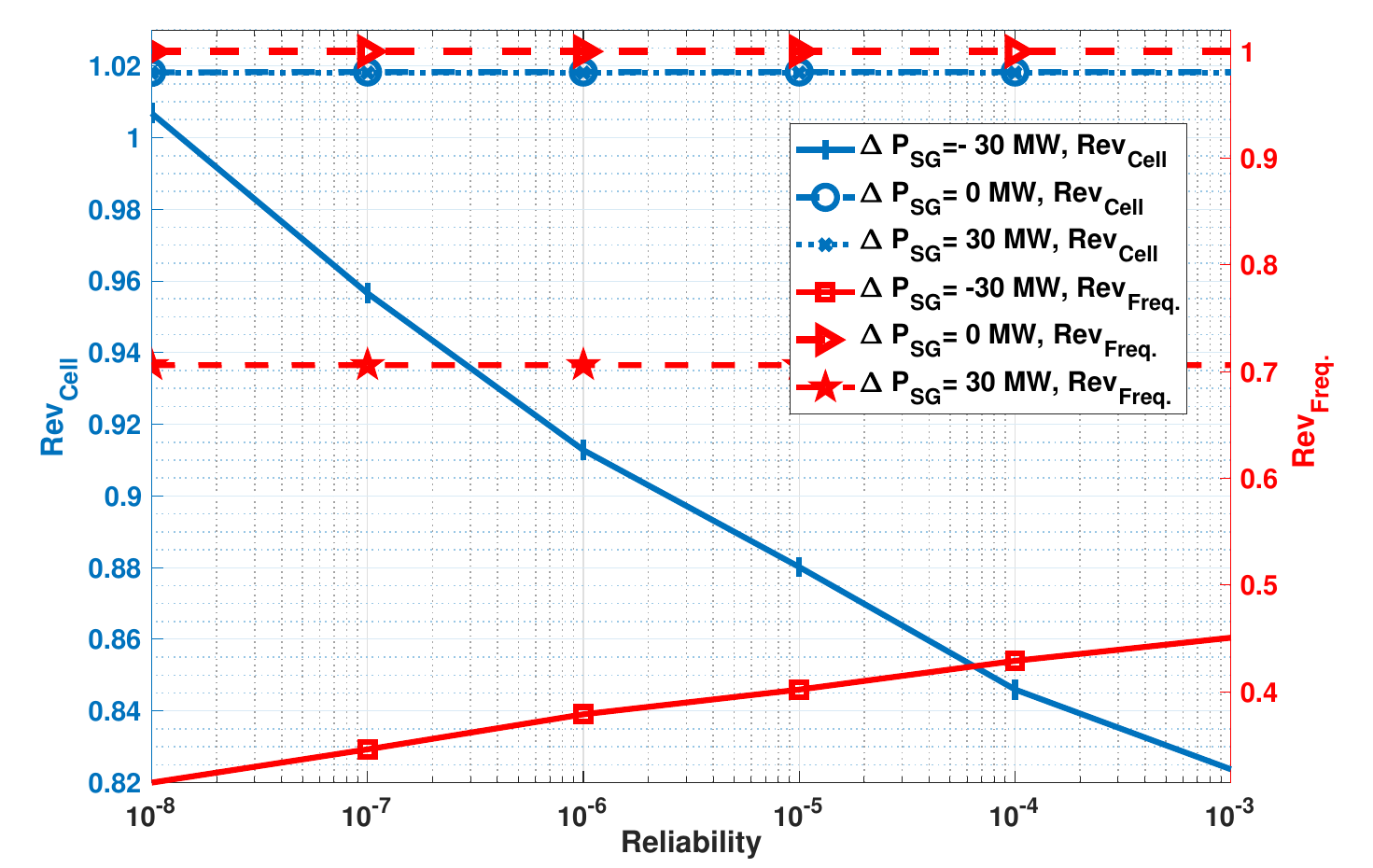}
    \caption{Reliability Impact on Frequency Regulation in Smart Grid}
    \vspace{-0.5cm}
    \label{fig-sim-Reliability}
\end{figure}

Here, we investigate the impacts of URLLC communication KPIs such as reliability and delay on frequency regulation within the smart grid. As depicted in Fig. \ref{fig-sim-Reliability}, an increasing in the reliability requirements for URLLC users leads to a reduction in power vacancy compensation through cellular network,i.e., $\text{Rev}_\text{Freq.}$ decreases. The figure indicates that for a negative power vacancy of $-30$ MW,  each unit increase in reliability results in $3\%$ reduction in $\text{Rev}_\text{Freq.}$ for a network with $1500$ BSs. By relaxing the reliability constraint, $\text{Rev}_\text{Cell}$ decreases because a lower data rate is required, leading to reduced energy consumption for communication.  This energy conservation can be redirected back to the smart grid to regulate frequency, consequently enhancing the revenue of frequency regulation of the smart grid through the cellular network BSs.

\begin{figure}[t]
    \centering
    \includegraphics[width=0.93\linewidth]{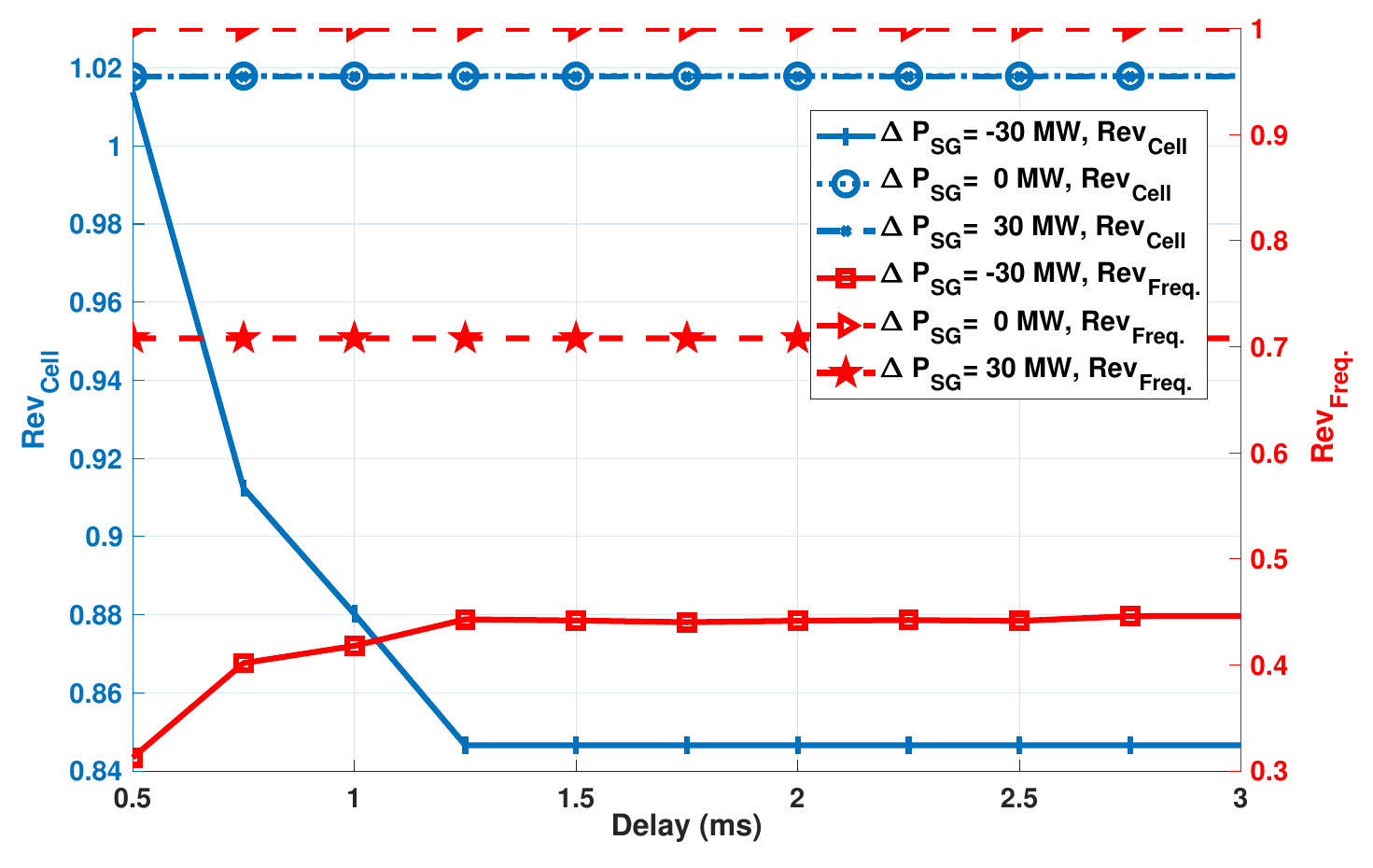}
    \caption{Delay Impact on Frequency Regulation in Smart Grid}
    \vspace{-0.5cm}
    \label{fig-sim-Delay}
\end{figure}

Similarly, in Fig \ref{fig-sim-Delay}, the impact of delay on frequency regulation within the smart grid, under a fixed reliability of $99.99999$, is demonstrated. As observed, for a power vacancy of $-30$ MW, increasing the delay to $1.25$ ms, increases the revenue of smart grid frequency regulation to $0.46$, after which it remains constant. This is because, for  delay exceeding $1.25$ ms, the reliability constraint becomes dominant, and further increases in delay do not affect the optimization problem. 

\begin{figure}[t]
    \centering
    \includegraphics[width=0.88\linewidth]{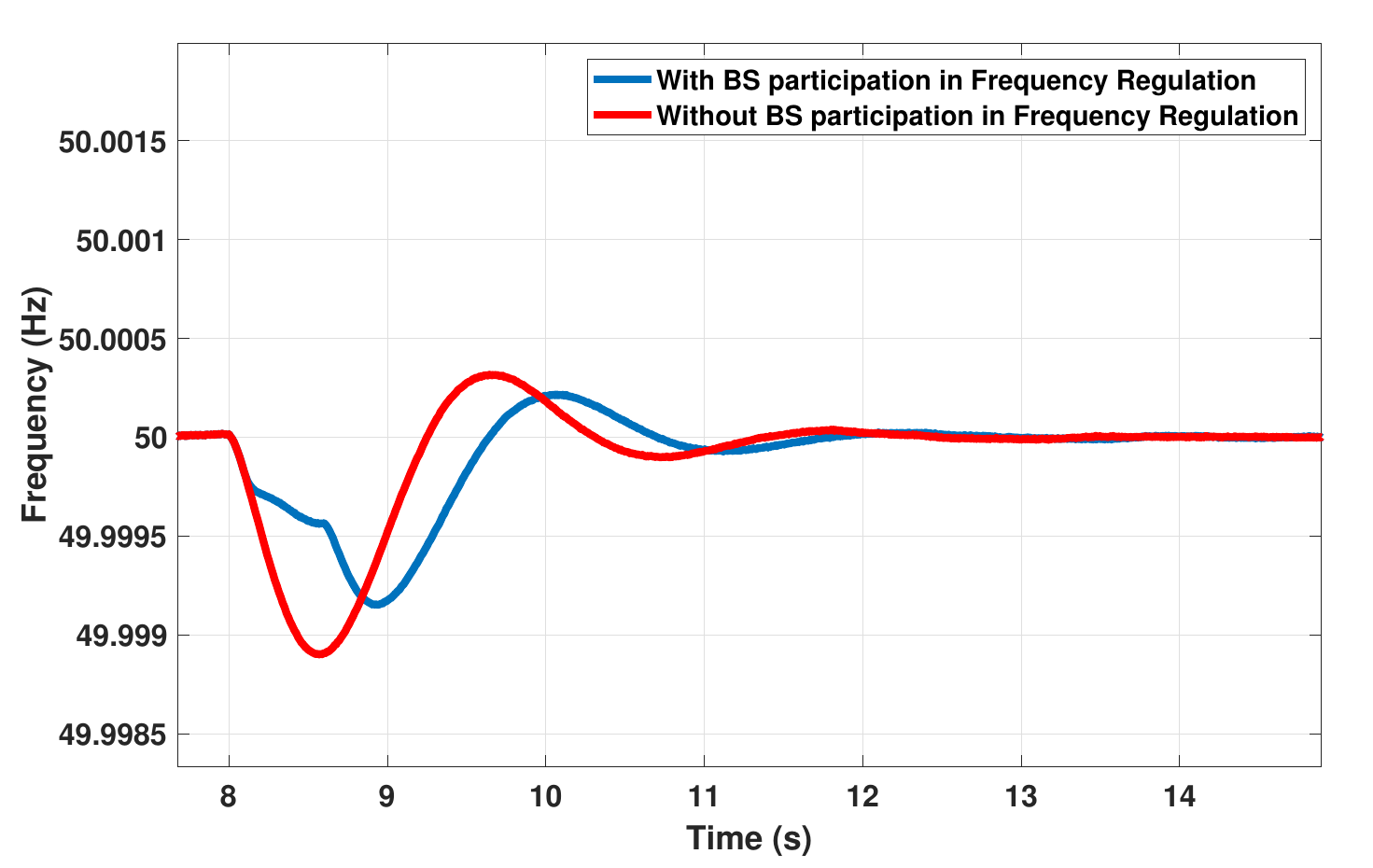}
    \caption{Frequency regulation in $\Delta P_{SG}=30 MW$}
    \vspace{-0.5cm}
    \label{SimFigFreq1}
\end{figure}

By considering this frequency regulation done by the BSs, the transient behaviour of frequency is shown in Figure \ref{SimFigFreq1}. Comparing the frequency with BS participation by the operation of the system without frequency regulation of the BSs, shows that the overshoot of the system has been reduced about 26\%  and the system is tend to get stabilized in a lower time.

In the case of zero power vacancy, the participation of cellular network in frequency regulation and hence $\text{Rev}_\text{Freq.}$ is almost equal to $1$. This means the summation of injected/delivered energy to/from the smart grid is zero, and the communication network relies on its own stored energy in the battery. The stored energy in the battery is sufficient for the communication network to operate at least $5$ hours with full load. Therefore, due to the objective function which aims to maximize the total rate, the BSs operate at full load, maintaining the total rate and hence communication service revenue $\text{Rev}_\text{Cell}$ at its highest value.
For positive power vacancy of $30$ MW, the communication  network utilizes smart grid electricity to fully charge its batteries and uses the excess energy from the smart grid to maximize communication service revenue. In this scenario, the communication network operates at full load, maintaining the highest data rate and $\text{Rev}_\text{Cell}$, and the participation in the smart grid for $1500$ BSs remains at $0.7$. Consequently $70$\% of the extra energy of the smart grid can be used through cellular network. 

\begin{figure}[t]
    \centering
    \includegraphics[width=0.88\linewidth]{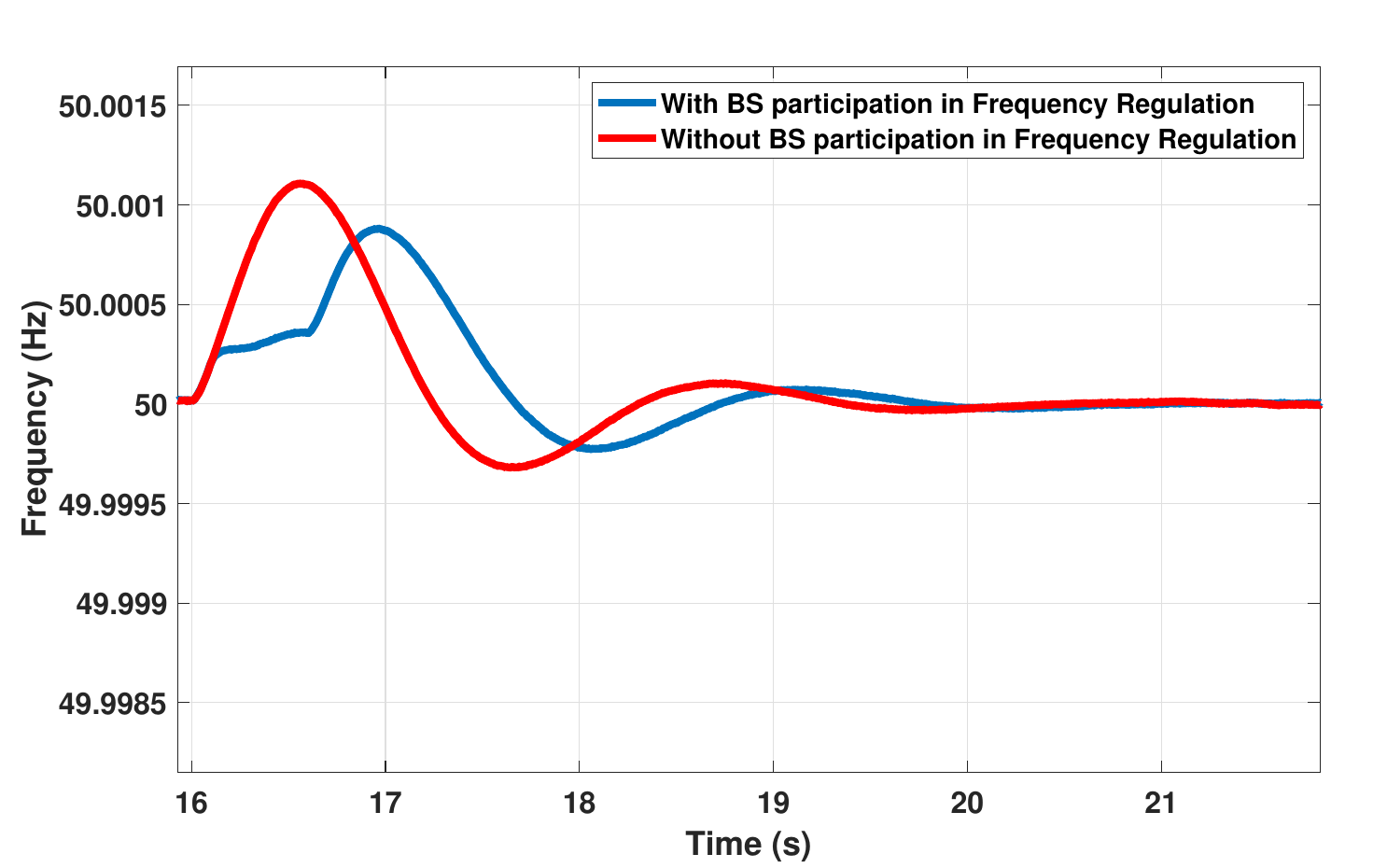}
    \caption{Frequency regulation in $\Delta P_{SG}=-30 MW$}
    \vspace{-0.5cm}
    \label{SimFigFreq2}
\end{figure}

In this mode, the transient behaviour of frequency is shown in Figure \ref{SimFigFreq2} and by comparing the between participating of BSs and not participating of BSs, it is seen that overshoot of system's frequency regulation has been reduced about 30\%. 

\vspace{-0.3cm}
\subsection{Comparison to Disjoint Approach}
For the evaluation of our proposed system model, we compare our approach with a scenario in which the communication network and smart grid are treated independently. Consequently, we consider two separate optimization problems: one is solved within the communication network, and the other within the smart grid independently. In this disjoint approach, the smart grid first optimizes revenue of smart grid frequency regulation  by considering a fixed load in the communication network, i.e., the worst-case scenario in which the communication network is operating at peak load. Subsequently, the smart grid solves the following optimization problem:
	\begin{align}\label{optproblem}
	&\mathop {\max }\limits_{
		\scriptstyle{\bf{Z}},{ \bf{\hat E}},{\bf{K}}} \text{Rev}_\text{Freq.}^{t},
	\\\text{s.t.:}&\text{~(C8)-(C11)}, \nonumber
	\end{align}
On the other hand, the communication network maximizes its revenue of communication service  with a fixed power vacancy compensation value, which is determined based on the previous optimization problem. Therefore, we have the following optimization problem:
\begin{align}\label{optproblem}
	&\mathop {\max }\limits_{\scriptstyle{\bf{ P}},{\boldsymbol{\rho}}, {\bf{D}}\hfill\atop
		\scriptstyle{\bf{}}{ \bf{}}{\bf{}}\hfill} \text{Rev}_\text{Cell}^{t} 
	\\\text{s.t.:}&\text{~(C1)-(C9)}, \nonumber
	\end{align}

\begin{figure}[t]
    \centering
    \includegraphics[width=0.93\linewidth]{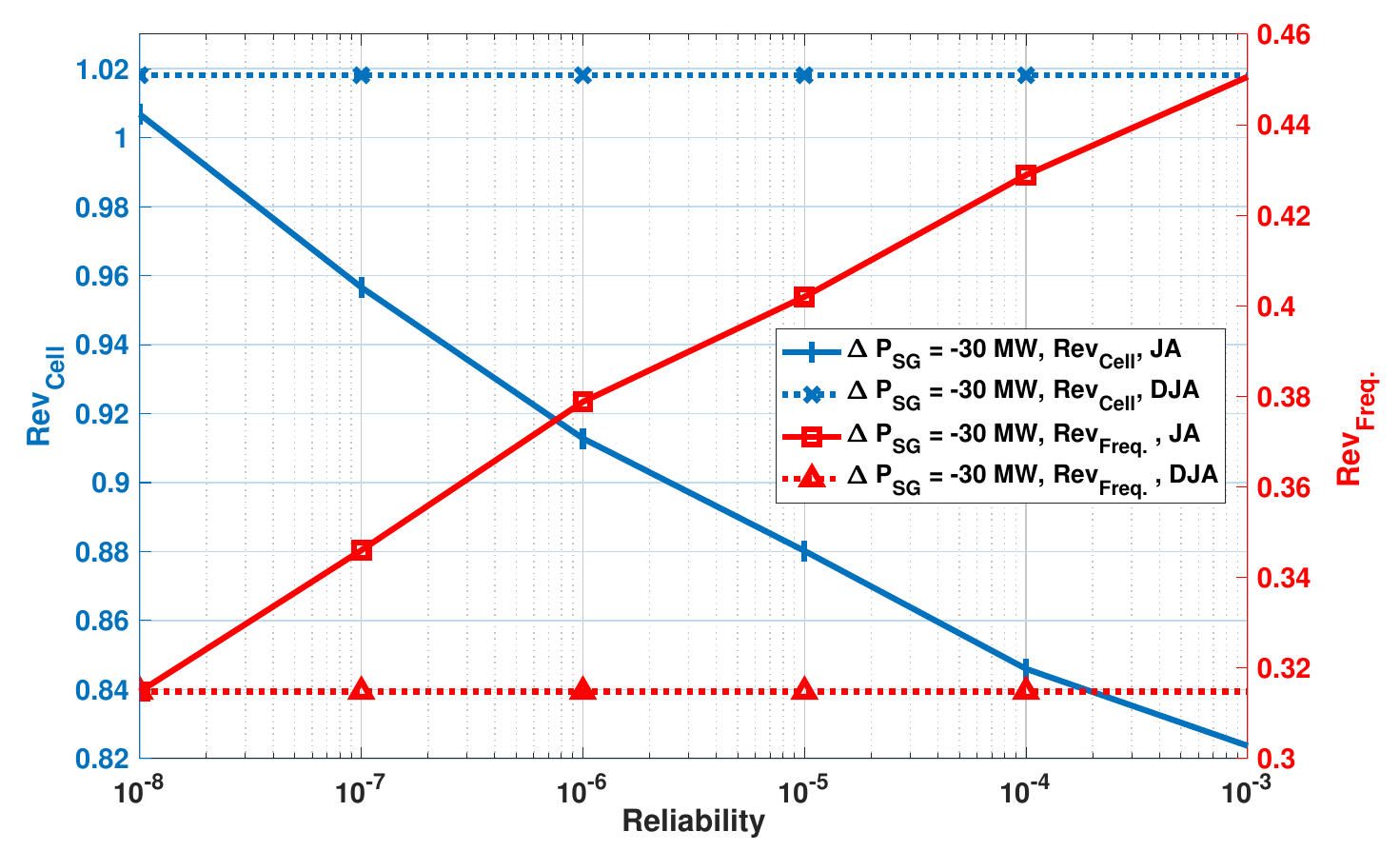}
    \caption{Frequency regulation: Joint Approach in Comparison to Disjoint Approach}
    \vspace{-0.5cm}
    \label{SimFigFreq44}
\end{figure}

As can be seen in Fig. \ref{SimFigFreq44}, in the disjoint approach, for negative power vacancy, by considering the worst-case scenario in the communication network, the revenue of smart grid frequency regulation remains fixed at the lowest value, and on the other hand, the revenue of communication service remains fixed at the maximum value. In the joint approach, these values become optimized based on the conditions of the smart grid and communication network jointly, which leads to regulate frequency in the smart grid and satisfying communication requirements jointly.
\vspace{-0.3cm}
\section{Conclusion}
This paper has demonstrated the potential of cellular BS energy storages to enhance smart grid resilience and efficiency by participating in the energy flexibility market. By allocating a portion of their capacity to support smart grid frequency regulation services, BS storages can provide significant benefits without compromising communication user requirements such as delay and reliability. The novel joint resource allocation problem formulated in this study maximizes both the total sum rate in the communication network and the participation of BSs in frequency regulation, while considering battery aging and cycling constraints.
Simulation results indicate that cellular networks can significantly enhance power vacancy compensation by adjusting user reliability and delay requirements. Specifically, a network with $1500$ BSs can increase power vacancy compensation from $31$\% to $46$\% by reducing reliability from $10^{-8}$ to $10^{-3}$ or increasing delay from $0.5$ ms to $3$ ms. This translates to an increase in power vacancy compensation from $9.3$ MW to $13.5$ MW for a power vacancy of $-30$ MW, surpassing the capacity of a wind turbine.
Overall, the integration of BS energy storages into smart grid operations presents a promising approach for sustainable energy management and cost-effectiveness for cellular network operators. Future work will focus on further optimizing the resource allocation strategies and exploring additional applications of BS storages in smart grid environments.

\balance
\bibliographystyle{IEEEtran}
\bibliography{Mybib}

\begin{thebibliography}{10}
\providecommand{\url}[1]{#1}
\csname url@samestyle\endcsname
\providecommand{\newblock}{\relax}
\providecommand{\bibinfo}[2]{#2}
\providecommand{\BIBentrySTDinterwordspacing}{\spaceskip=0pt\relax}
\providecommand{\BIBentryALTinterwordstretchfactor}{4}
\providecommand{\BIBentryALTinterwordspacing}{\spaceskip=\fontdimen2\font plus
\BIBentryALTinterwordstretchfactor\fontdimen3\font minus
  \fontdimen4\font\relax}
\providecommand{\BIBforeignlanguage}[2]{{%
\expandafter\ifx\csname l@#1\endcsname\relax
\typeout{** WARNING: IEEEtran.bst: No hyphenation pattern has been}%
\typeout{** loaded for the language `#1'. Using the pattern for}%
\typeout{** the default language instead.}%
\else
\language=\csname l@#1\endcsname
\fi
#2}}
\providecommand{\BIBdecl}{\relax}
\BIBdecl

\bibitem{ITU-TechnicalSmart}
ITU-T, ``{Smart Energy Saving of 5G Base Station: Based on AI and other
  emerging technologies to forecast and optimize the management of 5G wireless
  network energy consumption},'' 2021.

\bibitem{SWGIMT2030}
S.~IMT-2030, ``{DRAFT NEW RECOMMENDATION ITU-R M.[IMT.FRAMEWORK FOR 2030 AND
  BEYOND], Framework and Overall Objectives of the Future Development of IMT
  for 2030 and Beyond},'' 2023.

\bibitem{6466419}
K.~Yang and A.~Walid, ``{Outage-Storage Tradeoff in Frequency Regulation for
  Smart Grid With Renewables},'' \emph{IEEE Transactions on Smart Grid},
  vol.~4, no.~1, pp. 245--252, 2013.

\bibitem{ding2023strategy}
G.~Ding, L.~Li, Y.~Li, Q.~Zhang, C.~Luo, Q.~Chen, and X.~Zheng, ``{Strategy of
  5G Base Station Energy Storage Participating in the Power System Frequency
  Regulation},'' \emph{Arabian Journal for Science and Engineering}, vol.~48,
  no.~11, pp. 14\,537--14\,548, 2023.

\bibitem{hosseini2022battery}
S.~A. Hosseini, M.~Toulabi, A.~Ashouri-Zadeh, and A.~M. Ranjbar, ``Battery
  energy storage systems and demand response applied to power system frequency
  control,'' \emph{International Journal of Electrical Power \& Energy
  Systems}, vol. 136, p. 107680, 2022.

\bibitem{arrigo2020assessment}
F.~Arrigo, E.~Bompard, M.~Merlo, and F.~Milano, ``Assessment of primary
  frequency control through battery energy storage systems,''
  \emph{International Journal of Electrical Power \& Energy Systems}, vol. 115,
  p. 105428, 2020.

\bibitem{suarez2012overview}
L.~Suarez, L.~Nuaymi, and J.-M. Bonnin, ``{An Overview and Classification of
  Research Approaches in Green Wireless Networks},'' \emph{Eurasip journal on
  wireless communications and networking}, vol. 2012, pp. 1--18, 2012.

\bibitem{9260159}
A.~El~Amine, H.~A.~H. Hassan, and L.~Nuaymi, ``{Battery-Aware Green Cellular
  Networks Fed By Smart Grid and Renewable Energy},'' \emph{IEEE Transactions
  on Network and Service Management}, vol.~18, no.~2, pp. 2181--2192, 2021.

\bibitem{6831472}
J.~Leithon, S.~Sun, and T.~J. Lim, ``{Energy Management Strategies for Base
  Stations Powered by the Smart Grid},'' in \emph{2013 IEEE Global
  Communications Conference (GLOBECOM)}, 2013, pp. 2635--2640.

\bibitem{6883952}
J.~Leithon, T.~J. Lim, and S.~Sun, ``{Energy exchange among base stations in a
  Cellular Network through the Smart Grid},'' in \emph{2014 IEEE International
  Conference on Communications (ICC)}, 2014, pp. 4036--4041.

\bibitem{6687957}
------, ``{Online Energy Management Strategies for Base Stations Powered by the
  Smart Grid},'' in \emph{2013 IEEE International Conference on Smart Grid
  Communications (SmartGridComm)}, 2013, pp. 199--204.

\bibitem{heylen2021challenges}
E.~Heylen, F.~Teng, and G.~Strbac, ``{Challenges and Opportunities of Inertia
  Estimation and Forecasting in Low-Inertia Power Systems},'' \emph{Renewable
  and Sustainable Energy Reviews}, vol. 147, p. 111176, 2021.

\bibitem{8399832}
C.~{She}, Z.~{Chen}, C.~{Yang}, T.~Q.~S. {Quek}, Y.~{Li}, and B.~{Vucetic},
  ``Improving network availability of ultra-reliable and low-latency
  communications with multi-connectivity,'' \emph{IEEE Transactions on
  Communications}, vol.~66, no.~11, pp. 5482--5496, Nov 2018.

\bibitem{8253477}
C.~{She}, C.~{Yang}, and T.~Q.~S. {Quek}, ``Joint uplink and downlink resource
  configuration for ultra-reliable and low-latency communications,'' \emph{IEEE
  Transactions on Communications}, vol.~66, no.~5, pp. 2266--2280, May 2018.

\bibitem{7529226}
G.~Durisi, T.~Koch, and P.~Popovski, ``{Toward Massive, Ultrareliable, and
  Low-Latency Wireless Communication With Short Packets},'' \emph{Proceedings
  of the IEEE}, vol. 104, no.~9, pp. 1711--1726, 2016.

\bibitem{8541123}
C.~Sun, C.~She, C.~Yang, T.~Q.~S. Quek, Y.~Li, and B.~Vucetic, ``{Optimizing
  Resource Allocation in the Short Blocklength Regime for Ultra-Reliable and
  Low-Latency Communications},'' \emph{IEEE Transactions on Wireless
  Communications}, vol.~18, no.~1, pp. 402--415, 2019.

\bibitem{9512400}
I.~Muhammad, H.~Alves, N.~H. Mahmood, O.~L.~A. López, and M.~Latva-aho,
  ``{Mission Effective Capacity—A Novel Dependability Metric: A Study Case of
  Multiconnectivity-Enabled URLLC for IIoT},'' \emph{IEEE Transactions on
  Industrial Informatics}, vol.~18, no.~6, pp. 4180--4188, 2022.

\bibitem{9322582}
Y.~Wu, D.~Qiao, and H.~Qian, ``{Efficient Bandwidth Allocation for URLLC in
  Frequency-Selective Fading Channels},'' in \emph{GLOBECOM 2020 - 2020 IEEE
  Global Communications Conference}, 2020, pp. 1--6.

\bibitem{1210731}
D.~Wu and R.~Negi, ``{Effective capacity: a wireless link model for support of
  quality of service},'' \emph{IEEE Transactions on Wireless Communications},
  vol.~2, no.~4, pp. 630--643, 2003.

\bibitem{8638940imp}
C.~Guo, L.~Liang, and G.~Y. Li, ``{Resource Allocation for Low-Latency
  Vehicular Communications: An Effective Capacity Perspective},'' \emph{IEEE
  Journal on Selected Areas in Communications}, vol.~37, no.~4, pp. 905--917,
  2019.

\bibitem{lorincz2012measurements}
J.~Lorincz, T.~Garma, and G.~Petrovic, ``{Measurements and Modelling of Base
  Station Power Consumption Under Real Traffic Loads},'' \emph{Sensors},
  vol.~12, no.~4, pp. 4281--4310, 2012.

\bibitem{ETSIBSPower}
ETSI, ``{ETSI TS 102 706-2 V1.5.1 (2018-11), Environmental Engineering (EE);
  Metrics and Measurement Method for Energy Efficiency of Wireless Access
  Network Equipment; Part 2: Energy Efficiency - dynamic measurement method
  },'' 2018.

\bibitem{ITU-TPBS}
ITU-T, ``{L.1390, Energy saving technologies and best practices for 5G radio
  access network (RAN) equipment, Energy efficiency, smart energy and green
  data centres},'' 2018.

\bibitem{9204689}
I.~A. Umoren, M.~Z. Shakir, and H.~Tabassum, ``{Resource Efficient
  Vehicle-to-Grid (V2G) Communication Systems for Electric Vehicle Enabled
  Microgrids},'' \emph{IEEE Transactions on Intelligent Transportation
  Systems}, vol.~22, no.~7, pp. 4171--4180, 2021.

\bibitem{6678362}
D.~T. Ngo, S.~Khakurel, and T.~Le-Ngoc, ``J{oint Subchannel Assignment and
  Power Allocation for OFDMA Femtocell Networks},'' \emph{IEEE Transactions on
  Wireless Communications}, vol.~13, no.~1, pp. 342--355, 2014.

\bibitem{9146520}
N.~Gholipoor, S.~Parsaeefard, M.~R. Javan, N.~Mokari, H.~Saeedi, and
  H.~Pishro-Nik, ``{Resource Management and Admission Control for Tactile
  Internet in Next Generation of Radio Access Network},'' \emph{IEEE Access},
  vol.~8, pp. 136\,261--136\,277, 2020.

\bibitem{8686217}
M.~Moltafet, S.~Parsaeefard, M.~R. Javan, and N.~Mokari, ``{Robust Radio
  Resource Allocation in MISO-SCMA Assisted C-RAN in 5G Networks},'' \emph{IEEE
  Transactions on Vehicular Technology}, vol.~68, no.~6, pp. 5758--5768, 2019.

\bibitem{grant2014cvx}
M.~Grant and S.~Boyd, ``{CVX: Matlab Software for Disciplined Convex
  Programming, Version 2.1},'' 2014.

\bibitem{ngo2014joint}
D.~T. Ngo, S.~Khakurel, and T.~Le-Ngoc, ``Joint subchannel assignment and power
  allocation for {OFDMA} femtocell networks,'' \emph{IEEE Transactions on
  Wireless Communications}, vol.~13, no.~1, pp. 342--355, January 2014.

\bibitem{8070468}
C.~She, C.~Yang, and T.~Q.~S. Quek, ``{Cross-Layer Optimization for
  Ultra-Reliable and Low-Latency Radio Access Networks},'' \emph{IEEE
  Transactions on Wireless Communications}, vol.~17, no.~1, pp. 127--141, 2018.

\bibitem{7841709}
------, ``{Cross-Layer Transmission Design for Tactile Internet},'' in
  \emph{2016 IEEE Global Communications Conference (GLOBECOM)}, 2016, pp. 1--6.

\bibitem{9851476}
A.~Mansour~Saatloo, A.~Mehrabi, M.~Marzband, and N.~Aslam, ``{Hierarchical
  User-Driven Trajectory Planning and Charging Scheduling of Autonomous
  Electric Vehicles},'' \emph{IEEE Transactions on Transportation
  Electrification}, vol.~9, no.~1, pp. 1736--1749, 2023.

\end{thebibliography}
\end{document}